\documentclass[journal,comsoc]{IEEEtran}
\usepackage{lipsum}% http://ctan.org/pkg/lipsum
\usepackage{setspace}
\IEEEoverridecommandlockouts
\usepackage{amsmath}
\usepackage{commath}
\usepackage{multirow}
\usepackage{color}
\usepackage{float}
\usepackage{cite}
\usepackage[utf8]{inputenc}
\usepackage{lmodern}
\usepackage{csquotes}
\usepackage[bottom]{footmisc}
\usepackage[justification=centering]{caption}
\usepackage{blindtext}
\usepackage{wrapfig}
\usepackage{graphicx}
\usepackage{amssymb}
\usepackage{amsthm}
\usepackage{array}\usepackage{algorithm,algorithmic}
\usepackage{booktabs}
\usepackage{graphicx}
\usepackage{epstopdf}\usepackage[utf8]{inputenc}
\usepackage[english]{babel}
\DeclareGraphicsExtensions{.eps,.pdf,.gif,.png,.jpg}
\usepackage{verbatim} 
\usepackage[hyphens,spaces,obeyspaces]{url}
\usepackage{lineno,nohyperref}
\usepackage{epstopdf} %converting to PDF
\modulolinenumbers[5]

\newtheorem{definition}{Definition}
\RequirePackage{filecontents}
\usepackage{ragged2e}
\usepackage{url} 
\usepackage[T1]{fontenc}
\usepackage[utf8]{inputenc}
\usepackage{authblk}
\usepackage{microtype} %remove empy space in URL

\newcommand{\vect}[1]{\boldsymbol{#1}}
\newcommand*{\email}[1]{#1}
\def\BibTeX{{\rm B\kern-.05em{\sc i\kern-.025em b}\kern-.08em
		T\kern-.1667em\lower.7ex\hbox{E}\kern-.125emX}}
\makeatletter
\def\moverlay{\mathpalette\mov@rlay}
\def\mov@rlay#1#2{\leavevmode\vtop{%
		\baselineskip\z@skip \lineskiplimit-\maxdimen
		\ialign{\hfil$\m@th#1##$\hfil\cr#2\crcr}}}
\newcommand{\charfusion}[3][\mathord]{
	#1{\ifx#1\mathop\vphantom{#2}\fi
		\mathpalette\mov@rlay{#2\cr#3}
	}
	\ifx#1\mathop\expandafter\displaylimits\fi}
\makeatother
\newcommand*{\defeq}{\mathrel{\vcenter{\baselineskip0.5ex \lineskiplimit0pt
			\hbox{\scriptsize.}\hbox{\scriptsize.}}}%
	=}

\newtheorem{remark}{Remark}

\begin{document}	
	\title{
Deep Learning Based Caching for Self-Driving Cars in Multi-access Edge Computing}
	\author{Anselme~Ndikumana,~Nguyen~H.~Tran,~\IEEEmembership{Senior~Member,~IEEE,~} Do~Hyeon~Kim,~Ki~Tae~Kim,~and~Choong~Seon~Hong,~\IEEEmembership{Senior~Member,~IEEE}
		\thanks{Anselme Ndikumana is with the Faculty of Computing and Information Sciences, University of Lay Adventists of Kigali, KK 508 St, Kigali, Rwanda, and also with the Department of Computer Science and Engineering, Kyung Hee University, Yongin-si, Gyeonggi-do 17104,  Rep. of Korea, E-mail: \email{\{anselme\}@khu.ac.kr}}
		\thanks{Do Hyeon Kim, Ki Tae Kim, and Choong Seon Hong  are with the Department of Computer Science and Engineering, Kyung Hee University,  Yongin-si, Gyeonggi-do 17104, Rep. of Korea,\\  E-mail:\email{\{doma, glideslope, cshong\}@khu.ac.kr}}\\
		\thanks{Nguyen H. Tran is with the School of Computer Science, The University of Sydney, Sydney, NSW 2006, Australia, E-mail: \email{\{nguyen.tran\}@sydney.edu.au}\\\vspace*{-1cm}}}
	{}
	\maketitle

\begin{abstract}
Without steering wheel and driver’s seat, the self-driving cars will have new interior outlook and spaces that can be used for enhanced infotainment services. For traveling people, self-driving cars will be new places for engaging in infotainment services. Therefore, self-driving cars should determine themselves the infotainment contents that are likely to entertain their passengers. However, the choice of infotainment contents depends on passengers' features such as age, emotion, and gender. Also, retrieving infotainment contents at data center can hinder infotainment services due to high end-to-end delay. To address these challenges, we propose infotainment caching in self-driving cars, where caching decisions are based on passengers' features obtained using deep learning. First, we proposed deep learning models to predict the contents need to be cached in self-driving cars and close proximity of self-driving cars in multi-access edge computing servers attached to roadside units. Second, we proposed a communication model for retrieving infotainment contents to cache. Third, we proposed a caching model for retrieved contents. Fourth, we proposed a computation model for the cached contents, where cached contents can be served in different formats/qualities based on demands. Finally, we proposed an optimization problem whose goal is to link the proposed models into one optimization problem that minimizes the content downloading delay. To solve the formulated problem, a block successive majorization-minimization technique is applied. The simulation results show that the accuracy of prediction for the contents that need to be cached is $97.82\%$ and our approach can minimize the delay.
\end{abstract}

\IEEEpeerreviewmaketitle
\section{Introduction}
\label{sec:introduction}
\subsection{Background and Motivations} 

Recently, the automobile industries have focused on the next stage of autonomous driving, called \blockquote{self-driving}, where cars will drive themselves without human driver intervention \cite{daily2017self}. \textcolor{black}{To make the self-driving cars more intelligent, they need to be equipped with smart sensors and analytics tools that collect and analyze heterogeneous data related to passengers on-board, pedestrians, and the environment in real-time, in which Artificial Intelligence (AI) plays significant roles \cite{ferdowsi2017deep}.}  Furthermore, AI will be an empathetic companion of passengers for assisting them and providing personalized services. Therefore,  AI will need to understand passengers' features \cite{DrivingMarket}. 

In this work, we choose self-driving cars over human-driven cars because self-driving cars already have On-Board Units (OBUs) with  Graphics Processing Units (GPUs), Field Programmable Gate Array (FPGA), and Application Specific Integrated Chip (ASIC) to handle in-car AI. This gives the self-driving cars the capability to observe, think, learn, and navigate in real driving environments \cite{daily2017self}. Also, according to a study on the incremental time and what activities people will perform if everyone uses self-driving cars, it is estimated that there will be $22$ billions of hours for extra media consummation in the US \cite{Graham2018}. Therefore, with AI and OBUs that can handle Computation, Communication, Caching, and Control (4C) in self-driving cars, passengers will spend more time on infotainment services such watching media, playing games, and utilizing social networks. \textcolor{black}{ To support this, self-driving cars should be equipped with recent emerging technologies for infotainment services such as AI-based games, Virtual, Augmented, and Mixed Reality \cite{fathi2018big}.} However, retrieving infotainment contents from Data Centers (DCs) can worsen infotainment content delivery services due to the associated end-to-end delay and consumed backhaul bandwidth resource. As an example, watching a video in a car requires three components, namely a video source, screen, and sound system. Therefore, if the source of the video is not in the car, the car needs to download it from DC. Assuming the DC is distantly located, then the infotainment content delivery services will incur a high delay. Therefore, caching in self-driving cars will play an important role in enhancing infotainment services. Furthermore, for retrieving infotainment contents that need to be cached in self-driving cars, we consider Multi-access Edge Computing (MEC) \cite{hu2015mobile, ndikumana2018joint} as a suitable technology to support self-driving cars through caching infotainment contents near self-driving cars. In this work, MEC servers are deployed at RoadSide Units (RSUs).

\subsection{Challenges for infotainment Caching}

\begin{itemize}
	\item
	In human-driven cars, drivers choose the infotainment contents to display or play. However, in the absence of the driver, the self-driving car should determine itself the infotainment contents to cache and play that are likely to entertain its passengers.     
	\item Some infotainment contents may not be appropriate for consumption by passengers depending on their age and area. Therefore, the self-driving car should determine itself the infotainment contents to cache that do not violate prohibited and restricted content access policies.    
	\item As shown in Fig. \ref{Contentpreferences} generated from YouTube demographics dataset for one month available in \cite{NextAnalytics}, people have different content preferences, in which their choices depend on their features such as age and gender. Therefore, in the self-driving driving cars, caching decisions for the infotainment contents should depend on passengers' features.
	\item      
	Self-driving cars will eventually deliver more heterogeneous infotainment contents such as movies, TV, music, and games as well as recent emerging technologies such as Virtual, Augmented, and Mixed Reality \cite{fathi2018big}. However, obtaining infotainment contents from DC can induce high car-DC delay. Therefore, self-driving cars need to be supported by MEC servers by caching infotainment content in close proximity to self-driving cars at RSUs.    
	\item
	Self-driving cars are sensitive to delay due to their high mobility and connection in-motion. Therefore, to achieve less variation in transmission delay for downloading contents need to be cached, at the beginning of the journey, the self-driving car should select available MEC servers en-route that will be used to download infotainment contents. 
\end{itemize}

\subsection{Related Works} 

\textcolor{black}{Content caching at macro Base Stations (BSs) and RSUs has gained significant attention \cite{ndikumana2017collaborative, ndikumana2018joint}.} However, there is still a lack of literature on caching infotainment contents in the cars based on passengers' features. To address the above challenges, in \cite{divine2013auto}, the author proposed an auto-control system for the vehicle infotainment system, where the system analyzes the characteristics of passengers, e.g., by listening to conversations between passengers, understanding the atmosphere or ambiance inside the vehicle during the trip, and determining the relationship between passengers. The results of this analysis help the system identify and deliver appropriate infotainment contents to the passengers. However, in \cite{divine2013auto}, there is no caching approach for infotainment contents. Always the cars have to retrieve the infotainment contents from DC. In \cite{ma2017low}, the authors proposed a cloud-based vehicular ad-hoc network, where both vehicles and  RSUs participate in content caching. However, introducing a cloud-based controller into vehicle caching can increase the content retrieval delay. \textcolor{black}{ To overcome this issue, the authors in \cite{kazmi2019infotainment} proposed joint communication, caching, and computation. However, the authors did not discuss how to select the contents to cache based on vehicle occupants. Furthermore, for V2X communication, authors in \cite{varanasi2019adaptive} proposed the caching approach which is based on machine learning, where they used different classes of data and class-based cache replacement schemes.} Other alternatives have been proposed in \cite{yuan2018toward}, where the authors considered two levels of caching at the edge servers (BSs) and autonomous cars. In their proposal, the edge servers inject contents into some selected cars that have enough influence to share these contents with other cars. However, in a realistic network environment, BSs and cars may belong to different entities. Therefore, without an incentive mechanism, there is no motivation for car owners to allow BS operator(s) to inject contents into their cars and participate in content sharing. Finally, in \cite{raichelgauz2018system}, the authors proposed a method for caching in an autonomous car. In their proposal, autonomous vehicles have cache storages to cache the data collected by the sensors, including metadata related to driving decisions. From the cache storage, it is possible to generate a driving decision based on similar previous cached driving decisions.

\subsection{Contributions} 

To address the aforementioned challenges, we propose a deep learning based caching for self-driving cars, where caching decisions depend on passengers' features obtained using deep learning approaches and available communication, caching, and computation (3C) resources. As an extended version of our earlier work published in \cite{ndikumana2019self}, the main contributions of this paper are summarized as follows:
\begin{figure}[t]
	\centering
	\includegraphics[width=0.95\columnwidth]{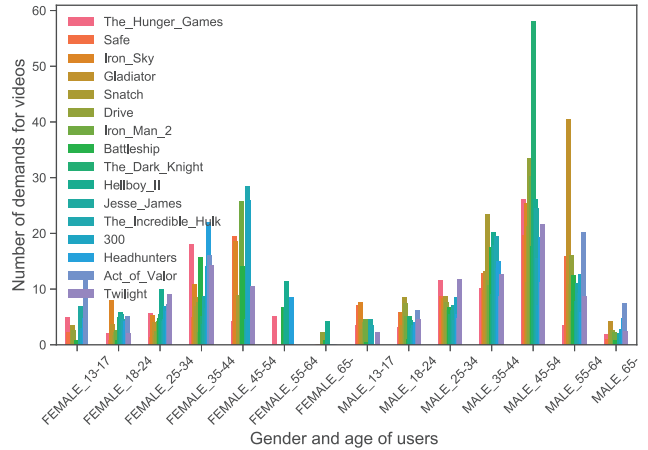}
	\caption{Content preferences based on users' features\cite{NextAnalytics}.}
	\label{Contentpreferences}
\end{figure}
\begin{itemize}
	\item 
	\textcolor{black}{We propose deep learning based caching for self-driving cars as a new application of Convolutional Neural Network (CNN), where caching decisions depend on passengers' features obtained using CNN model and facial images of the passengers.} Here, we assume the CNN model is trained and tested at DC using dataset. Then, the CNN model is deployed at MEC servers attached to the RSUs in close proximity to the self-driving cars, where the self-driving cars can retrieve model with minimized delay.
	\item      
	\textcolor{black}{We propose a Multi-Layer Perceptron (MLP) framework at DC to predict the probability of infotainment contents to be requested in specific edge areas of MEC servers. Then, the MLP prediction output is deployed at MEC servers. During off-peak hours, each MEC server uses MLP output to identify the infotainment contents that have high predicted probability values of being requested in its area, downloads and caches them. To identify the infotainment contents that are likely to entertain its passengers and need to be cached in the self-driving car, each self-driving car downloads and stores the CNN model and MLP output from the MEC server. The self-driving car uses the CNN model for predicting passengers' features via facial images captured by its camera. Then,  the self-driving car compares the CNN output with the MLP output using classification \cite{whang2015non, martineau2009improving} for identifying the contents that meet passengers' features. }
	\item
	\textcolor{black}{We propose a communication model that helps the self-driving car select available RSUs en-route. Then, the self-driving car uses these RSUs for retrieving identified infotainment contents that meet passengers' features and need to be cached.}
	\item
	\textcolor{black}{We propose a computation model for cached infotainment contents, where the cached contents can be served in different formats and qualities depending on demands. Therefore, we consider that MEC servers and self-driving cars have computation resources, which can be used to compute or process cached contents in different formats and qualities.}
	\item
	\textcolor{black}{We formulate an optimization problem that links the formulated models (deep learning-based caching, communication, and computation models) into one optimization problem whose goal is to minimize the content downloading delay. However, the formulated problem is shown to be non-convex. Therefore, to make it convex, we proposed a convex surrogate problem, which is an upper-bound of the formulated problem.  Then, we apply the Block Successive Majorization-Minimization (BS-MM) technique \cite{sun2017majorization} for solving it. We chose  BS-MM over other optimization techniques because BS-MM is a new technique that can decompose the original problem  into small subproblems, where each subproblem can be solved separately.}
\end{itemize}

Specifically, the novelties of our proposal over the related works in \cite{ndikumana2018joint, zhang2017cost, hu2017roadside, chen2017distribution, ndikumana2018joint2, chang2018learn, he2018integrated, ma2017low, yuan2018toward} are as follows: To the best of our knowledge, we are the first to investigate self-driving car caching for infotainment contents, where caching decisions are based on passengers' features and available communication, caching, and computation resources. 

% Paper structure 
The rest of the paper is organized as follows.  We discuss the system model in Section \ref{sec:system-model} and present our deep learning based caching approach in Section \ref{sec:CachingInSelfDrivingCar}. In Section \ref{sec:Problem_solution}, we discuss the problem formulation and solution. We present a performance evaluation in Section \ref{sec:PerformanceEvaluation}. Finally, we conclude the paper in Section \ref{sec:Conclusion}. 

\section{System model}
\label{sec:system-model}
%---------------------------------------------------------------
\begin{figure}[t]
	\centering
	\includegraphics[width=1.0\columnwidth]{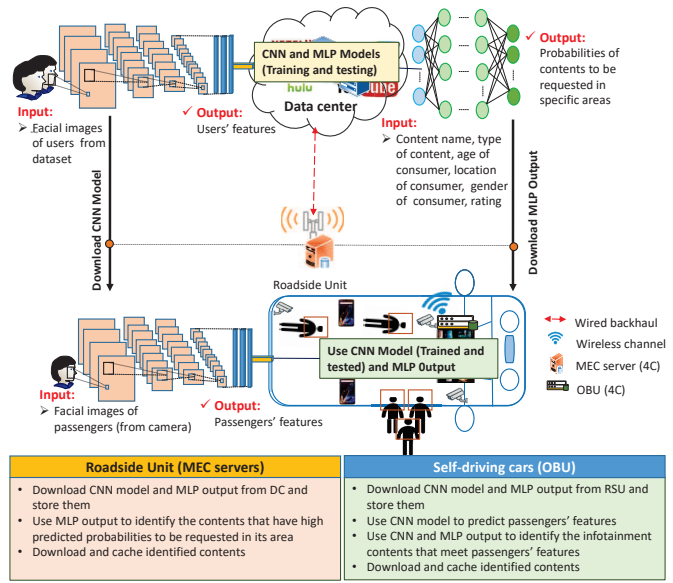}
	\caption{\textcolor{black}{Illustration of our system model.}}
	\label{fig:SystemModel}
\end{figure}
The system model of deep learning based caching is depicted in Fig. \ref{fig:SystemModel}.

\emph{Data Center (DC):} We assume that DC has higher computation resources than the self-driving car and RSU. Therefore, to minimize computation time, \textcolor{black}{ we use DC and dataset to make, train, and test deep learning models (CNN  and  MLP models) that will be used for predicting passengers features and infotainment contents need to be cached at the RSUs and in self-driving cars}. To reduce the communication delay between the self-driving cars and the DC, the trained and tested CNN model and  MLP output are deployed at MEC servers attached to the RSUs.

\emph{RoadSide Unit (RSU):} As defined in 3GPP TS 22.185 V15.0.0 \cite{V2X}, we consider eNB-type RSU as an entity that supports both evolved NodeB (eNB) functionalities and V2X applications. We assume that each RSU $r \in \mathcal{R}$ has access to the DC via a wired backhaul of capacity $\omega_{r, DC}$, where  $\mathcal{R}$ is the set of RSUs. Also, each RSU $r \in \mathcal{R}$ has an MEC server. Therefore, unless stated otherwise, we use the terms “RSU” and “MEC server” interchangeably. Furthermore, as defined in  3GPP specifications in \cite{V2X}, we consider an MEC server as locally application server that serves a certain particular geographic area  $n \in \mathcal{N}$, where $\mathcal{N}=\{1, 2, \dots, N\}$ is a set of geographic areas. Furthermore, each MEC server $ r \in \mathcal{R}$ has a cache storage of capacity $c_r$ and computational resource of capacity $p_r$.  Furthermore, during off-peak hours, by using backhaul communication resources, each RSU $r \in \mathcal{R}$ downloads CNN model and  MLP output. Then, based on the MLP output, each MEC server downloads and cache infotainment contents that have high predicted probabilities of being requested in its area. We use $\mathcal{I}$ to denote a set of infotainment contents,  where each content $ i \in \mathcal{I}$ has a size of $S(i)$ Mb. Also, we consider that people from different areas may need different infotainment contents \cite{blaszczyszyn2015optimal}. Therefore, it is more reasonable to cache infotainment contents at RSUs based on probabilities of being requested in particular areas.

\emph{Self-driving car:} We consider $\mathcal{V}$ as a set of self-driving cars, where each self-driving car $ v \in \mathcal{V}$ has OBU that can handle 4C to support caching and computation of infotainment contents for passengers.  Furthermore, each self-driving $ v \in \mathcal{V}$ can get broadband Internet service from RSU $r \in \mathcal{R}$  through a wireless link of capacity $\omega_{v, r}$.  Each self-driving car $ v \in \mathcal{V}$ has a cache storage capacity of $c_v$ and computation capability of $p_v$. Furthermore, to predict the passengers' features, we use the CNN model. This helps in deciding which infotainment contents to request and cache in the self-driving car that meet passengers' features. During off-peak hours, each self-driving car $ v \in \mathcal{V}$ downloads CNN model and MLP output from MEC server. By using the k-means and binary classification,  the self-driving car compares its CNN prediction with the predicted output from MLP. This helps the self-driving car identify the infotainment contents that are appropriate to the passengers' features. Finally, the self-driving car downloads and caches the identified contents that meet passengers' features. 
	
To avoid repetitive delivery of the same contents that require to use backhaul bandwidth, depending on demands, we consider that the computation resources of RSU and the self-driving car can be used to compute cached infotainment contents. As an example, content $i'$ with the H.264 format may not be available in the cache storage. Instead, the cache storage may have content $i$ with the MP4 format of the same content. Therefore, to satisfy the demand, by using the computational resource, cached infotainment content $i$ can be converted to content $i'$ (MP4 to H.264).

\begin{table}[t]
	\caption{Summary of key notations.}
	\label{tab:table1}
	\begin{tabular}{ll}
		\toprule
		Notation & Definition\\
		\midrule
		$\mathcal{R}$ & Set of RSUs, $|\mathcal{R}|= R$\\
		$\mathcal{V}$ &Set of self-driving cars,  $|\mathcal{V}|= V$\\
		$\mathcal{I}$ & Set of contents, $|\mathcal{I}|= I$\\
		$\mathcal{I}_r(n)$ & Set of contents that need to be cached\\ 
		& in area $n$  of RSU $r$, $|\mathcal{I}_r(n)|= I_r(n)$\\ 
		$\mathcal{U}$ &Set of consumers of contents,  $|\mathcal{U}|= U$\\
		$\vect{x}$ &Input of MLP\\
		$\vect{\tilde{y}}$ &Output of MLP\\
		$\vect{y}$ &Ground truth for MLP\\
		$M$ & The number of input features\\
		$N$ & The number of geographic areas\\
		$c_r$ & Caching capacity of each  RSU	$r \in \mathcal{R}$\\
		$p_r$ & Computation capability of   RSU	$r \in \mathcal{R}$\\
		$c_v$ & Caching capacity of each  car	$v \in \mathcal{V}$ \\
		$p_v$ &  Computation capability of  car $v \in \mathcal{V}$\\
		$\tau^\textrm{Tot}_{u}(\vect{q}, \vect{h}, \vect{\varrho})$ & Total delay experienced by each\\
		& passenger $u\in \mathcal{U}_v$\\ 
		$\psi^v_u$ & Data rate for each passenger $u$ via WiFi \\
		& of self-driving car $v$\\ 
		\bottomrule
	\end{tabular}
\end{table}
%---------------------------------------------------------------
\section{Deep Learning Based Caching in Self-Driving Cars}
\label{sec:CachingInSelfDrivingCar}

In this section, to identify the infotainment contents need to be cached, we discuss the deep learning and recommendation model in Section \ref{subsec:ComparisonModel}. For retrieving the recommended contents requires communication resources. Therefore, in Section \ref{subsec:CommunicationModel}, we discuss the communication model. For caching downloaded contents, we present the caching model in Section \ref{subsec:caching_model}. Furthermore, Based on the demands,  cached contents can be converted or transcoded to different formats by using computational resources, where the computation model is described in Section \ref{subsec:computation_model}.

\subsection{Deep Learning and Recommendation Model}
\label{subsec:ComparisonModel} 
In this subsection, we discuss MLP  for predicting infotainment contents need to be cached at RSUs nearby the self-driving cars, CNN model for predicting passengers' features, and recommendation model for identifying the contents that meet passengers' features and need to be cache in the self-driving cars. 

\subsubsection{ Multi-Layer Perceptron (MLP)}
\label{subsubsec:MLP} 
We propose  MLP for predicting probabilities of contents to be requested in particular areas of RSUs. \textcolor{black}{We choose MLP over other prediction methods such as AutoRegressive (AR) and AutoRegressive Moving Average (ARMA) models because MLP can cope with both linear and non-linear prediction problems \cite{azzouni2018neutm}}. We use a demographical dataset that will be described in Section \ref{sec:PerformanceEvaluation}. The input and output are described as follows: 
\begin{itemize}
	\item
	\emph{Input:} In the dataset, we have infotainment content names, rating, viewer's age, gender, and location as the input of MLP. Furthermore, for predicting the probabilities of contents to be requested in specific areas,   we use $\vect{x}=(x_1,x_2, \dots x_M)^T$ to denote the input vector, where the subscripts are used to denote the features such as content names, rating, viewer's age, gender, and location.
	\item
	\emph{Output:} From the input, MLP tries to predict $\vect{\tilde{y}}=(\tilde{y}_1,\tilde{y}_2, \dots \tilde{y}_N)^T$ as the output vector and the subscripts are used to denote the geographic areas.  Also, in the output layer, each area $n \in \mathcal{N}$ corresponds to one neuron, where the output layer predicts the probabilities of contents to be cached in each specific area  $n \in \mathcal{N}$.
\end{itemize}

For MLP, we use $l$ to denote the  number of hidden layers, $\vect{x}$ for the input vector, $\vect{b}^{(1)}, \dots, \vect{b}^{(l)}$ for the bias vectors,  $\vect{W}^{(1)}, \dots, \vect{W}^{(l)}$ for the weight matrices at each hidden layer, and $\vect{\tilde{y}}$ for  the output vector. $\vect{\tilde{y}}$ can be expressed as follows:
\begin{equation}
\label{eq:weigted_sum_MLP}
\begin{aligned}
\vect{\tilde{y}}= f(\vect{W}^{(l)} \dots f(\vect{W}^{(2)}f(\vect{W}^{(1)}\vect{x}+\vect{b}^{(1)})+\vect{b}^{(2)}) \dots +\vect{b}^{(l)}).
\end{aligned}
\end{equation}
where $f(.)$ is the activation function.

In our MLP, we use the Rectified Linear Unit (ReLU) as the activation function in all the layers except at the output layer. We chose ReLU over other activation functions, because it mitigates the vanishing gradient problem  experienced by MLP during the training process \cite{alom2018history}. Furthermore, in the output layer $l$, we use the softmax function as an activation function.  The purpose of the softmax function is to squeeze the output vector $\vect{\tilde{y}}$ into a set of probability values, where softmax function is defined as: 
\begin{equation}
\label{eq:softmax}
\begin{aligned}
softmax(\vect{\tilde{y}})^{(l)}=\frac{e^{\tilde{y}_l}}{\sum\nolimits_{n=1}^{N} e^{\tilde{y}_n}},\;  \text{for $l = 1,\dots, N$}.
\end{aligned}
\end{equation}
The output layer has $N$ neurons that correspond to $N$ areas of RSUs. Furthermore, for the error function, we chose the cross-entropy  error function over other error functions since our MLP classifies the contents needs to be cached in $N$ geographic areas of RSUs.  This problem can be considered as a classification problem, where we interpret the output as probabilities of the contents to be requested in each specific area $n \in \mathcal{N}$. The cross-entropy error function $A(\vect{y}, \vect{\tilde{y}})$ can be expressed as follows:
\begin{equation}
A(\vect{y}, \vect{\tilde{y}}) =- \sum\nolimits_{n=1}^{N}y_n\log \tilde{y}_n.
\end{equation}
$A(\vect{y}, \vect{\tilde{y}})$ calculates the cross-entropy between the estimated class probabilities 
$\vect{\tilde{y}}$ and the ground truth $\vect{y}$. 

Finally, to reduce the communication delay between the self-driving car and DC, as the DC may be located far from the self-driving cars,  the output of the MLP are downloaded and stored to the RSUs based on their areas.

\subsubsection{Convolutional Neural Network (CNN)}
\label{subsubsec:CNN}
In our proposal, we do not focus on proposing new CNN model. Conversely, we focus on a new application of existing CNN model for automatic age, emotion, and gender prediction from facial images \cite{dehghan2017dager} in caching decision. We describe the CNN workflow for automatic age, emotion, and gender extraction as follows:
\begin{itemize}
	
	\item
	\emph{Input:} We consider $\vect{k}_0$ as the input image with three-dimensional space: height, width, and the number of color channels (red, green, and blue). 
	\item
	\emph{Convolution layer:} The convolution layer applies filters to the input regions and computes the output of each neuron. Each neuron is connected to local regions of the input, and using dot products between the weight and local regions, the convolution layer produces a feature map $\vect{k}_j$. We use $\vect{k}_j$ to denote the feature map produced after convolution layer $j$.
	\item
	\emph{RELU layer:} In this layer, we apply the ReLU ($\max(0,\vect{k}_j)$) as an elementwise activation function. The ReLU keeps the size of its associated convolution layer $j$ unchanged.
	\item
	\emph{Max pooling layer:}  After the convolution and RELU layers, we have a high-dimensional matrix. Therefore, for dimension reduction, we apply a max-pooling layer as a downsampling operation.
	\item
	\emph{Fully-connected layer:} This layer is fully connected to all previous neurons and is used to compute the class scores that a face could potentially belong to. Here, we have two classes for gender (male and female),  $101$ classes for age (from $0$ to $101$), and $8$ classes for emotion (anger, anticipation, disgust, fear, joy, sad, surprise, and trust). In other words, we use three fully-connected layers for age, gender, and emotion classification.
	\item
	\emph{Softmax layer and output:} In this layer, for each facial image, we need to interpret the output as the probability values of classes for gender, emotion, and age that a facial image could potentially belong to. To achieve this, the softmax activation function is applied to the output of the fully-connected layers.
\end{itemize}

\textcolor{black}{To reduce the communication delay between the self-driving cars and DC, the trained and tested CNN model is deployed to the RSUs. Then, each self-driving car $v \in \mathcal{V}$ downloads CNN model and uses it for predicting age, gender, and emotion of passengers from facial images. Once the facial image of a passenger is captured via a camera. The self-driving car can extract features such as eyes, nose, mouth, and chin and use them for classifying the passengers' faces into different age, emotion, and gender classes.  As describe in below recommendation model, this helps the self-driving car identify the infotainment contents that meet  passengers' features as recommended contents to cache.} \textcolor{black}{Here, we assume that the passengers are aware of the presence of the camera. In other words, the self-driving cars have warning signs that inform passengers on the presence of the cameras. The same techniques were used in the deployment of public video surveillance at streets or  public places \cite{van2016privacy}.}

\begin{figure}[t]
	\centering
	\includegraphics[width=0.90\columnwidth]{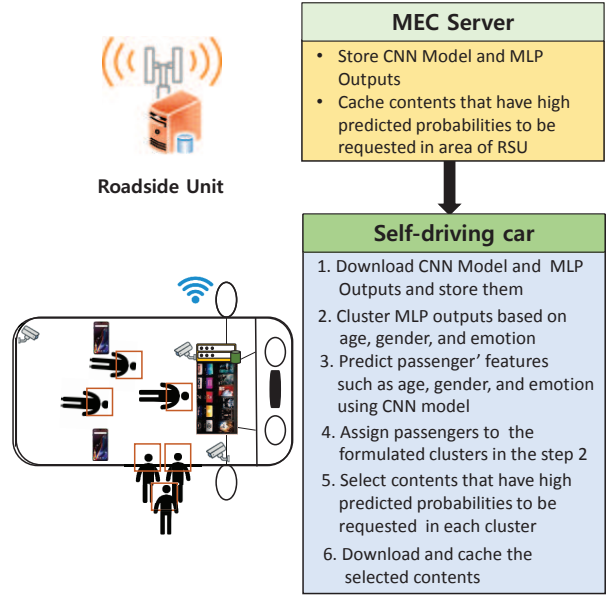}
	\caption{Recommendation model for self-driving car.}
	\label{fig:recommendation_model}
\end{figure}

\subsubsection{Recommendation Model}
\label{subsubsec:Recommendation}
The workflow of the recommendation model for self-driving cars is illustrated in Fig. \ref{fig:recommendation_model} and described as follows:

\begin{itemize}
	\item
	\emph{Step 1:} Each self-driving car  $v \in \mathcal{V}$ downloads the MLP output and CNN model from MEC server attached to RSU.
	\item
	\emph{Step 2:} By using the k-means algorithm for age and emotion-based grouping and binary classification for gender-based grouping on the MLP output, each self-driving car $v \in \mathcal{V}$ creates age, gender, and emotion-based clusters of content consumers and generates an initial recommendation for the contents that need to be cached and have high  predicted probability values for being requested.         
	\item
	\emph{Step 3:} For each new passenger $u \in \mathcal{U}$, the self-driving car uses the CNN model for predicting its age, gender, and emotion from facial image.
	\item 
	\emph{Step 4:} The self-driving car uses these passenger's features to calculate the similarity of passenger $u\in \mathcal{U}$  with the existing users (i.e., content consumers) in age, gender, and emotion-based clusters. Then, based on the similarity calculation, each passenger $u \in \mathcal{U}$  will be assigned to a cluster.
	\item 
	\emph{Step 5:} After clustering the passenger(s), self-driving car $v \in \mathcal{V}$  selects top contents that have high predicted probability values for being requested as recommended contents to cache.
	\item 
	\emph{Step 6:} Finally, self-driving car $v \in \mathcal{V}$ downloads the recommended contents via  RSU and caches them in its cache storage $c_v$.
\end{itemize}

 For the k-means algorithm, first, we use age as numerical data. We denote $ \vect{\tilde{y}}_n$ as the MLP output at each area $n \in \mathcal{N}$ and  $\mathcal{X}=\vect{\tilde{y}}_n$  as the input of the k-means algorithm. The k-means partitions the consumer of the contents $\mathcal{X}=\{x_1, \dots, x_U \}$  into $K$  age-based clusters  $\mathcal{X}_1,\dots, \mathcal{X}_K$ such that $\mathcal{X}_1\cup \mathcal{X}_2 \cup\dots\cup \mathcal{X}_K=\mathcal{X}$. In k-means, consumers are grouped into clusters based on their age. In addition, the clusters are disjoint $\mathcal{X}_i \cap \mathcal{X}_j=\emptyset, \;  i\neq j$. The goal of k-means is to assign users to age-based clusters such  that the  objective function below is minimized:
\begin{equation}
\begin{aligned}
\underset{\{\mathcal{X}_j\}^K_{j=1}}{\text{min}} \;\sum_{j=1}^K   \sum_{x_u \in \mathcal{X}_j}\lVert x_u-\tilde{x}_j\rVert^2,
\end{aligned}
\label{eq:kmean}
\end{equation}
where $\tilde{x}_j$ is the centroid of cluster $\mathcal{X}_j$, which is defined as
\begin{equation}
\tilde{x}_j  =\frac{\sum_{x_u \in \mathcal{X}_j}x_u}{|\mathcal{X}_j|}.
\end{equation}

In addition to age, consumers in the same age-based cluster can have different choice for contents based on emotion. Therefore, in each age-based cluster $j$, we use the k-means algorithm to class the consumers of contents in $E$ emotion-based clusters (fear, sad, neutral, angry, disgusted, surprised). Therefore, in each emotion-based cluster $e$, we group users based on gender. For gender-based grouping, we  apply binary classification as described in \cite{martineau2009improving}, which results in the formation of two groups, one group for females (denoted $\mathcal{G}_{je}^\textrm{female}$) and another group for males (denoted $\mathcal{G}_{je}^\textrm{male}$) such that $\mathcal{G}_{je}^\textrm{female} \cap \mathcal{G}_{je}^\textrm{male}=\emptyset$. Then, inside  $\mathcal{G}_{je}^\textrm{female}$ and $ \mathcal{G}_{je}^\textrm{male}$  clusters, which are sub-clusters of age-based cluster $j$ and emotion-based cluster $e$, the self-driving car select top  infotainment contents that have  high predicted probability values of being requested as the recommended contents to cache. Finally, the self-driving car downloads and caches recommended infotainment contents.

In this work, we assume that the self-driving cars and MEC servers download and store the CNN model and MLP output during off-peak hours. Therefore, hereafter, we only focus on recommended infotainment contents downloading, caching, and computing.
\enlargethispage{2\baselineskip} 
\subsection{Communication Model for Retrieving Contents}
\label{subsec:CommunicationModel} 
\begin{figure}[t]
	\centering
	\includegraphics[width=0.90\columnwidth]{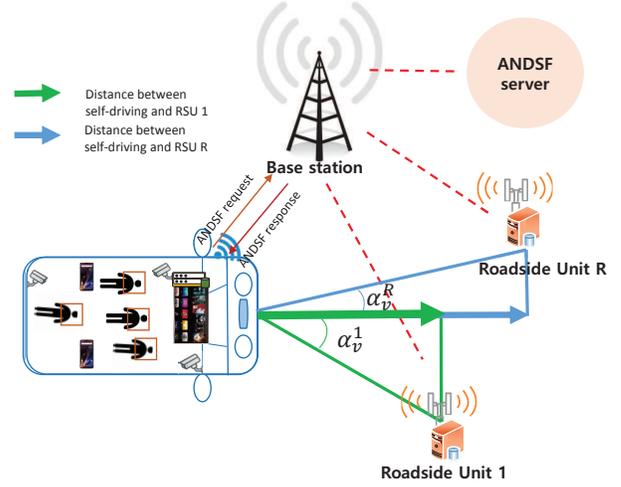}
	\caption{Communication planning for self-driving car.}
	\label{fig:communication_model}
\end{figure}
Using a backhaul link of capacity $\omega_{r, DC}$, each MEC server downloads the infotainment contents that have high predicted probability values for being requested in its area $n\in \mathcal{N}$. The transmission delay for downloading contents from the DC to the MEC server $r$ is:
\begin{equation}
\setlength{\jot}{10pt}
\tau^\textrm{DC}_r =
\frac{q^{\textrm{DC}\rightarrow r}\sum_{i \in \mathcal{I}_r(n)} S(i)}{\omega_{r, DC}}, 
\end{equation}
where  $\mathcal{I}_r(n)$ for $n \in \mathcal{N}$ denotes the set of predicted contents that have high probability values for being requested in area $n$ of RSU, and $q^{\textrm{DC}\rightarrow r}$ is a decision variable that indicates whether or not MEC server $r$  is connected to the the DC, such that: 
\begin{equation}
\setlength{\jot}{10pt}
q^{\textrm{DC}\rightarrow r}=
\begin{cases}
1,\; \text{if MEC server $r$ is connected to the DC,} \\
0,\;\text{otherwise.}
\end{cases}
\end{equation}

As illustrated in Fig. \ref{fig:communication_model}, to have less variation in the transmission delay and hand-off before the self-driving car starts its journey, it can select RSUs that will be used to download the top-recommended contents. To discover RSUs located in a route of the self-driving car, Access Network Discovery and Selection Function (ANDSF) implemented in the cellular network and described in 3GPP TS 24.312 V15.0.0 \cite{ANDSF} can be utilized. We assume each self-driving car $v \in \mathcal{V}$ moves in an area covered by macro Base Stations (BSs) and RSUs. Therefore, to obtain RSU information such as coordinate and coverage, the self-driving car sends a request to the ANDSF server via a BS \cite{ndashimye2016novel}. The request includes a  geographic location of the self-driving car, speed, and direction. On the other hand, in the ANDSF server’s feedback includes the coordinates and coverage of all RSUs available in the direction of the self-driving car.

Each self-driving car $v$ computes the following distance $\tilde{d}^r_v$ between each RSU $r$ and its route:
\begin{equation}
\label{eq:RSU-road}
\begin{aligned}
\tilde{d}^r_v= g^r_v sin \alpha^r_v,
\end{aligned}
\end{equation}
where $\alpha^r_v$ is the angle between the trajectory of movement of self-driving car $v$ and the straight line from RSU $r \in \mathcal{R}$, and  $g^r_v$ is the geographical distance between self-driving car $v$ and cache-enabled RSU $r$. In addition, each self-driving car $v$ computes the following distance $d^v_r$ remaining to reach each area covered by cache-enabled RSU $r \in \mathcal{R}$:
\begin{equation}
\label{eq:distanceCar_RSU}
\begin{aligned}
d^v_r= g^r_v cos \alpha^r_v.
\end{aligned}
\end{equation}
We defined $\rho^r_v$ as the probability that RSU $r \in \mathcal{R}$ will be selected as a source of infotainment contents to be cached in self-driving car $v$ as follows:
\begin{equation}
\setlength{\jot}{10pt}
\rho^r_v=
\begin{cases}
1,\; \text{if $\tilde{d}^r_v=0$}, \\
\frac{\tilde{d}^r_v}{\gamma_r}\;\text{if $0<\tilde{d}^r_v<\gamma_r$,}\\
0,\;\text{otherwise,}
\label{eq:probability_RSU}
\end{cases}
\end{equation}
where $\gamma_r$ is the radius of the area covered by  RSU $r \in \mathcal{R}$. Therefore, we define $q^r_v$ as a decision variable that indicates whether or not the self-driving car is connected to RSU $r \in \mathcal{R}$ as follows: 
\begin{equation}
\setlength{\jot}{10pt}
q^r_v =
\begin{cases}
1,\; \text{if $\rho^r_v>0$ and $d^v_r=0$,} \\
0,\;\text{otherwise.}
\label{eq:probability_RSU_variable}
\end{cases}
\end{equation}
Equations (\ref{eq:probability_RSU}) and (\ref{eq:probability_RSU_variable}) ensure that once the self-driving car $v$  reaches an area covered by cache-enabled RSU $r \in \mathcal{R}$, it immediately starts downloading the recommended infotainment contents.

We assume each RSU $r$ has a wireless channel of capacity $\omega_{v, r}$, where $\omega_{v, r}$ can be expressed as follows:  
\begin{equation}
\label{eq:capacity-RSU}
\begin{aligned}
\omega_{v, r}= q^r_v B_r \log_2\left(1 +  \varphi_r |G^r_v|^2\right),  \;\forall v \in \mathcal{V},\; r \in \mathcal{R},
\end{aligned}
\end{equation}
where $B_r$ is the bandwidth for the car to RSU communications, $G^r_v$ is the channel gain between RSU  $r$ and self-driving car $v$, and $\varphi_r$ is the transmission power of RSU  $r$. Therefore, based on the channel capacity, the transmission delay for downloading that meet passengers' features from the MEC server to self-driving car $v$ is expressed as: 
\begin{equation}
\setlength{\jot}{10pt}
\tau^r_{v}=\frac{\sum_{\tilde{i}_f, \tilde{i}_m \in \mathcal{I}_r(n)}q^r_v \left(
	S(\tilde{i}_f))+S(\tilde{i}_m)	\right)}{\omega_{v, r}},  
\end{equation}
where $\tilde{i}_f \in \mathcal{G}_{je}^\textrm{female}$ is the recommended infotainment content for female passengers and $\tilde{i}_m \in \mathcal{G}_{je}^\textrm{male}$ is the recommended infotainment content for male passengers in each age and emotion-based cluster in area $n$, where $\tilde{i}_f, \tilde{i}_m \in \mathcal{I}_r(n)$.

Based on self-driving car's speed, we consider $t^r_{v}$ as the time required by  self-driving car $v \in \mathcal{V}$ to leave an area covered by RSU $r$. We can calculate $t^r_{v}$ as follows:
\begin{equation}
\setlength{\jot}{10pt}
t^r_{v}=\frac{2q^r_v \gamma_r}{\mu_v},  
\end{equation}
where $\mu_v$ is the speed of self-driving car $v$. When $\tau^r_{v}< t^r_{v}$, the self-driving can easily download the recommended infotainment content in the area coverage by RSU $r$. However, when $\tau^r_{v} \geq  t^r_{v}$, the self-driving car can select the next RSU to use for downloading recommended infotainment contents.

Each self-driving car $v$ has a WiFi Router on board that can be used to provide WiFi connectivity to its passengers. However, in the self-driving car, passengers are free to choose their appropriate connections. Here, we aim to minimize delay experienced by the passengers that are inside of the self-driving car and use WiFi connectivity of the self-driving car for getting infotainment contents. Therefore, the instantaneous data rate for each passenger $u$ via the WiFi of self-driving car $v$ is given by:
\begin{equation}
\psi^v_u=\frac{q^v_u\varphi_v \tilde{\psi}^v_u \xi^v_u(|\mathcal{U}_v|)}{|\mathcal{U}_v|}, \forall u \in \mathcal{U}_v,\; v \in \mathcal{V}_v, 
\label{eq:instantaneous_data_(IWR)}
\end{equation} 
where $\varphi_v $ is the WiFi throughput efficiency factor and $|\mathcal{U}_v|$ is the number of passengers that are connected simultaneously to the WiFi of self-driving car $v$, where $\mathcal{U}_v \subset \mathcal{U}$. We use $\varphi_v $ to denote the overhead related to the MAC protocol layering. Furthermore, $\tilde{\psi}^v_u$ is the maximum theoretical data rate that the WiFi can handle. Furthermore, $\xi^v_u(|\mathcal{U}_v|)$ is a channel utilization function, which is a function of the number of passengers connected simultaneously to the WiFi \cite{cheng2016opportunistic}. $\xi^v_u(|\mathcal{U}_v|)$ is used to determine the impact of contention over the WiFi throughput. Also, we use $q^v_u$ as a decision variable that indicates whether or not passenger $u$ is connected to the WiFi of self-driving  $v$, specifically: 
\begin{equation}
\setlength{\jot}{10pt}
q^v_u=
\begin{cases}
1,\; \text{if the passenger $u$ is connected to the } \\
\; \; \; \;\text{WiFi of the self-driving car $v$,}\\
0,\;\text{otherwise.}
\end{cases}
\end{equation}
For each passenger $u \in \mathcal{U}_v$, based on its instantaneous data rate $\psi^v_u$, the transmission delay $\tau^v_{u}$ for downloading content $i$ from self-driving car $v$ is given by:
\begin{equation}
\setlength{\jot}{10pt}
\tau^v_{u}=\frac{\sum_{i \in \mathcal{I}_r(n)}q^v_u \left(
	S(\tilde{i}_f))+S(\tilde{i}_m)	\right)}{\psi^v_u}.  
\end{equation}

\subsection{ Caching Model for Retrieved Contents}
\label{subsec:caching_model} 

We assume that the cache storage $c_v$ of each self-driving car $v$ is limited. Therefore, the sizes of the recommended infotainment contents that need to be downloaded from the MEC server  and cached in the self-driving car must satisfy the cache resource constraint, which is expressed as follows:
\begin{equation}
\setlength{\jot}{10pt}
q^r_v\sum_{j=1}^K\left(\sum_{\tilde{i}_f \in \mathcal{G}_{je}^\textrm{female}}o_v^{\tilde{i}_f}S(\tilde{i}_f))  + \sum_{\tilde{i}_m \in \mathcal{G}_{je}^\textrm{male}}o_v^{\tilde{i}_m}S(\tilde{i}_m)\right) \leq c_v,
\end{equation}
where $o_v^{\tilde{i}_f}\in \{0,1\}$ is the decision variable that indicates whether or not self-driving car $v$ has to cache infotainment content $\tilde{i}_f \in \mathcal{G}_{je}^\textrm{female}$, where $o_v^{\tilde{i}_f}$ is given by:
\begin{equation}
\setlength{\jot}{10pt}
o_v^{\tilde{i}_f}=
\begin{cases}
1,\; \text{if self-driving car $v$ caches the content $\tilde{i}_f$},\\
0, \;\text{otherwise.}
\end{cases}
\end{equation}
On the other hand, we let $o_v^{\tilde{i}_m}\in \{0,1\}$ be the decision variable that indicates whether or not self-driving car $v$ has to cache infotainment content $\tilde{i}_m \in \mathcal{G}_{je}^\textrm{male}$, where $o_v^{\tilde{i}_m}$ is given by:
\begin{equation}
\setlength{\jot}{10pt}
o_v^{\tilde{i}_m}=
\begin{cases}
1,\; \text{if self-driving car $v$ caches the content $\tilde{i}_m$},\\
0, \;\text{otherwise.}
\end{cases}
\end{equation}
Furthermore, for analyzing cache storage utilization, which is based on cache hit and cache miss, we assume that $\tilde{i}_f$ and $\tilde{i}_m$ are cached in the same cache storage $c_v$. Therefore, we omit the subscript and superscript on content, and use $i$ to denote either content $\tilde{i}_f$ or $\tilde{i}_m$. 

We use  $h_i^{u\rightarrow v} \in \{  0, 1\}$  to denote the cache hit indicator at self-driving car $v$ for content $i \in \mathcal{I}_r(n)$ requested by customer $u \in \mathcal{U}$:
\begin{equation}
\setlength{\jot}{10pt}
h_{i}^{u\rightarrow v} =
\begin{cases}
1,\; \text{if content $i$ requested by consumer $u$ }\\
\; \; \;\text{ is returned from self-driving car $v$},\\
0, \;\text{otherwise.}
\end{cases}
\end{equation}

In the case of a cache miss ($h_i^{u\rightarrow v}=0$), the self-driving car needs to forward the demand for content $i$ to its associated MEC server. Based on the MLP output at the RSU, we assume that the MEC server caches the contents that have high probabilities of being requested in area $n$, where cache allocation has to satisfy the following constraint:
\begin{equation}
\label{eq:caching_constraint}
\begin{aligned}
q^{\textrm{DC}\rightarrow r}\sum_{i \in \mathcal{I}_r(n)}o_r^{i} S(i)\leq c_r,
\end{aligned}
\end{equation}
where  $o_r^{i}$ is a decision variable that indicates whether or not MEC server $r$ has to cache content $i \in \mathcal{I}_r(n)$, defined as follows:
\begin{equation}
\setlength{\jot}{10pt}
o_r^{i}=
\begin{cases}
1,\; \text{if MEC server $r$ caches content $ i \in \mathcal{I}_r(n)$,}\\
0, \;\text{otherwise.}
\end{cases}
\end{equation}
Furthermore, we use  $h_i^{r\rightarrow v} \in \{  0, 1\}$  to denote the cache hit indicator at the MEC server for content $i \in \mathcal{I}_r(n)$ requested by self-driving $v \in \mathcal{V}$:
\begin{equation}
\setlength{\jot}{10pt}
h_i^{r\rightarrow v} =
\begin{cases}
1,\; \text{if the content $ i$ requested by self-dring}\\
\; \; \; \;\text{ car $ v$ is cached at MEC server $r$} ,\\
0, \;\text{otherwise.}
\end{cases}
\end{equation}
However, when the MEC server does not have content $i$ in its cache storage, the MEC server forwards the demand for content $i$ to the DC via a wired backhaul link.

\subsection{Computation Model for Cached Contents}
\label{subsec:computation_model} 

In self-driving cars, a passenger may request a content format (e.g., H.264) that is not available in the cache storage $c_v$. Instead, the cache storage may have other content formats (e.g., MP4) for the same content that can be transcoded to the desired format (H.264). 

Therefore, to adopt this process of serving cached content after computation, we define the following decision variable: 
\begin{equation}
\setlength{\jot}{10pt}
h_{i'}^{v\rightarrow u} =
\begin{cases}
1,\; \text{if content $i'$ requested by consumer $u$ }\\
\; \; \;\text{ is returned by car $v$ after computation},\\
0, \;\text{otherwise.}
\end{cases}
\end{equation}

To ensure that self-driving car $v$  returns only one format of the requested content, the following constraint should be satisfied: 

\begin{equation}
\setlength{\jot}{10pt}
h_i^{u\rightarrow v} + h_{i'}^{v\rightarrow u} \leq  1. 
\end{equation}

We assume that converting content  $i$ to content $i'$ requires  computation resource $p^{i\rightarrow {i'}}_v$  of self-driving car $v$, where the computational resource allocation $p^{i\rightarrow {i'}}_v$ is given by:
\begin{equation}
p^{i\rightarrow {i'}}_v=p_v\frac{h_i^{u\rightarrow v}\varrho_{v}^{i\rightarrow{i'}}z^{i\rightarrow {i'}}}{\sum_{u \in \mathcal{U}}\sum_{i \in \mathcal{I}}h_i^{u\rightarrow v}\varrho_{v}^{i\rightarrow{i'}} z^{i\rightarrow {i'}}},\; \forall  v \in \mathcal{V},
\label{eq:computation_allocation_car}
\end{equation} where $z^{i\rightarrow {i'}}$ is the computation workload or intensity in terms of CPU cycles per bit required for converting cached content $i$ to $i'$, while $\varrho_{v}^{i\rightarrow{i'}}$ is the computation decision variable, which is expressed as:
\begin{equation}
\setlength{\jot}{10pt}
\varrho_{v}^{i\rightarrow{i'}} =
\begin{cases}
1,\; \text{if the cached content $i$ is converted to the }\\
\; \; \; \;\text{desired format $i'$ in self-driving car $v$}.\\
0, \;\text{otherwise.}
\end{cases}
\end{equation} 

In (\ref{eq:computation_allocation_car}), for computational resources allocation, we use weighted proportional allocation \cite{mosleh2016proportional} because it is simple to implement in practical communication systems such Vehicular Ad-hoc Networks (VANETs) and  4G \& 5G cellular networks \cite{ ndikumana2018joint}.  Furthermore, computation resource allocation should satisfy the following constraint:
\begin{equation}
\label{eq:computation_constraint_car}
\setlength{\jot}{10pt}
\sum_{u=1}^U\sum_{i=1}^{I_r(n)}q^v_u h_i^{u\rightarrow v}\varrho_{v}^{i\rightarrow{i'}} p^{i\rightarrow {i'}}_v \leq p_v.
\end{equation}
In addition, converting content $i$ to content $i'$ requires executing time. Therefore, in self-driving car $v$, the execution time $\tau^{i\rightarrow{i'}}_{v}$ is given by: 
\begin{equation}
\setlength{\jot}{10pt}
\tau^{i\rightarrow{i'}}_{v} =
\frac{q^v_u h_i^{u\rightarrow v}\varrho_{v}^{i\rightarrow{i'}}z^{i\rightarrow {i'}}  S(i)}{p^{i\rightarrow {i'}}_v}.
\end{equation}
When constraint (\ref{eq:computation_constraint_car}) cannot be satisfied due to insufficient computational resource for converting content  $i$ into the requested content $i'$, the self-driving car forwards the demand for content $i'$ to the MEC server.

At the MEC server, to convert cached content  $i$  into content $i'$, it requires an execution time of $\tau^{i\rightarrow{i'}}_r$. Thus, the execution time at the MEC server is given by:
\begin{equation}
\begin{aligned}
\setlength{\jot}{10pt}
\tau^{i\rightarrow{i'}}_r=(1-\varrho_{v}^{i\rightarrow{i'}})\left(
\frac{q^r_vh_i^{r\rightarrow v}\varrho_r^{i\rightarrow{i'}} z^{i\rightarrow {i'}}  S(i)}{p^{i\rightarrow {i'}}_r}\right),
\end{aligned}
\end{equation}
where $p^{i\rightarrow {i'}}_r $ is the required computation resource of MEC server $r$ for  converting content  $i$ to content $i'$. $p^{i\rightarrow {i'}}_r$ can be calculated in the same manner used in (\ref{eq:computation_allocation_car}). We define a $\varrho_r^{i\rightarrow{i'}}$  computation decision variable, where $\varrho_r^{i\rightarrow{i'}}$ is expressed as follows:
\begin{equation}
\setlength{\jot}{10pt}
\varrho_r^{i\rightarrow{i'}} =
\begin{cases}
1,\; \text{if the cached content $i$ is converted to }\\
\; \; \; \;\text{desired format $i'$  at MEC server} ,\\
0, \;\text{otherwise,}
\end{cases}
\end{equation}

We assume that the computation resources at the MEC  server are limited, where computation allocation has to satisfy the following constraint: 
\begin{equation}
\begin{aligned}
\label{eq:computation_constraint_mec_server}
\setlength{\jot}{10pt}
&\sum_{v=1}^V\sum_{i=1}^{I_r(n)}q^r_v h_i^{r\rightarrow v}\varrho_r^{i\rightarrow{i'}}p^{i\rightarrow {i'}}_r\leq P_r.
\end{aligned}
\end{equation}
In addition, we define $h_{i'}^{r\rightarrow v}$  as a decision variable that indicates whether or not the MEC server returns the requested content $i'$ to self-driving car $v$ after computation, where $h_{i'}^{r\rightarrow v}$ is given by:
\begin{equation}
\setlength{\jot}{10pt}
h_{i'}^{r\rightarrow v} =
\begin{cases}
1,\; \text{if content $i'$ requested by car $v$ is returned }\\
\; \; \;\text{  by MEC server $r$ after computation},\\
0, \;\text{otherwise.}
\end{cases}
\end{equation}

To ensure that converting  cached content $i$ to the requested content $i'$ is performed exactly at one location, either at the self-driving car or at MEC server, and self-driving car or  MEC server sends exactly one format of content, we formulate the following constraints: 
\begin{align} 
q^v_u(h_i^{u\rightarrow v} + h_{i'}^{v\rightarrow u})+q^r_v \eta_v(h_{i}^{r\rightarrow v}+h_{i'}^{r\rightarrow v}) &\leq 1, \\ 
\varrho_{v}^{i\rightarrow{i'}}+ q^r_v(1-\varrho_{v}^{i\rightarrow{i'}}) & \leq 1.
\end{align}
Here, we use $\eta_v=1- (h_{i}^{u\rightarrow v} + h_{i'}^{v\rightarrow u})$.
However, if the above constraints cannot be satisfied due to limited computation and caching resources, MEC server submits the request for content $i'$ to the DC.

\section{Problem Formulation and Solution}
\label{sec:Problem_solution}
In this section, we present our optimization problem for minimizing delay in downloading the infotainment contents in Section  \ref{subsec:problem_formulation}. Then, in Section \ref{subsec:ProposedSolution}, we present a solution of the formulated optimization problem.
\subsection{Problem Formulation}
\label{subsec:problem_formulation} 
In the self-driving car, to coordinate deep Learning \& recommendation, communication, caching, and computation models, we formulate an optimization problem that links the formulated models into one problem whose goal is to minimize total delay $\tau^\textrm{Tot}_{u}(\mathbf{q}, \mathbf{h}, \boldsymbol{\varrho})$ for retrieving infotainment contents, where  $\tau^\textrm{Tot}_{u}(\mathbf{q}, \mathbf{h}, \boldsymbol{\varrho})$ is given by: 
\begin{multline}
\label{eq:total_delay}
\setlength{\jot}{10pt}
\tau^\textrm{Tot}_{u}(\mathbf{q}, \mathbf{h}, \boldsymbol{\varrho}) =\tau^v_{u}h_{i}^{u\rightarrow v}+h_{i'}^{v\rightarrow u}\varrho_{v}^{i\rightarrow{i'}}\tau^{i\rightarrow{i'}}_{v} + \\ \left(1- (h_{i}^{u\rightarrow v}+\varrho_{v}^{i\rightarrow{i'}}h_{i'}^{v\rightarrow u})\right)(\tau^r_{v} h_i^{r\rightarrow v}+  \tau^{i\rightarrow{i'}}_r \varrho_{r}^{i\rightarrow{i'}}h_{i'}^{r\rightarrow v}) + \\(1- (h_i^{r\rightarrow v} +\varrho_{r}^{i\rightarrow{i'}}h_{i'}^{r\rightarrow v})) \tau^\textrm{DC}_r.
\end{multline}
In the above equation (\ref{eq:total_delay}), a requested infotainment content can be retrieved in the self-driving car. However, if the requested content can not be retrieved in a self-driving car, the self-driving car sends a request to RSU, where RSU can return the requested content. In the worst case, if the requested content can not be retrieved from self-driving car or RSU, DC can be used. Therefore, our optimization problem can be expressed as follows:
\begin{subequations}\label{eq:problem_formulation}
	\begin{align}
	&\underset{\vect{q}, \vect{h}, \vect{\varrho}}{\text{minimize}}\ \  \sum_{u=1}^U\tau^\textrm{Tot}_{u}(\vect{q}, \vect{h}, \vect{\varrho})
	\tag{\ref{eq:problem_formulation}}\\
	&\text{subject to:}\nonumber\\
	& \sum_{v=1}^Vq^r_v\leq 1, \;  \forall r \in \mathcal{R},
	\label{first:d}\\
	&q^r_v\sum_{j=1}^k(\sum_{\tilde{i}_f \in \mathcal{G}_{je}^\textrm{female}}o_v^{\tilde{i}_f}S(\tilde{i}_f))  + \sum_{\tilde{i}_m \in \mathcal{G}_{je}^\textrm{male}}o_v^{\tilde{i}_m}S(\tilde{i}_m)) \leq c_v, \label{first:e}\\
	&\sum_{u=1}^U\sum_{i=1}^{I_r(n)}q^v_u h_i^{u\rightarrow v}\varrho_{v}^{i\rightarrow{i'}} p^{i\rightarrow {i'}}_v \leq p_v,\;  \forall v \in \mathcal{V}, \;  \forall n\in \mathcal{N},
	\label{first:f}\\
	&q^v_u (h_i^{u\rightarrow v} + h_{i'}^{v\rightarrow u})+q^r_v \eta_v(h_{i}^{r\rightarrow v}+h_{i'}^{r\rightarrow v}) \leq  1,  \label{first:g}\\
	&q^v_u\varrho_{v}^{i\rightarrow{i'}}+ q^r_v(1-\varrho_{v}^{i\rightarrow{i'}})\leq 1.
	\label{first:h}\
	\end{align}
\end{subequations}

The constraint in (\ref{first:d}) ensures that the self-driving car has to be connected to RSU $r \in \mathcal{R}$ to download the contents. The constraints in (\ref{first:e}) and (\ref{first:f}) guarantee that the caching and computational resource allocations have to be less than or equal to the available caching and computational resources of the self-driving car.  Furthermore, constraint in (\ref{first:e}) is based on CNN output, where the self-driving car caches the contents based on passengers' features such as age, emotion, and gender.        
The constraint in (\ref{first:g}) ensures that the self-driving car or MEC server returns only one format of the requested content (either cached or computed from the cached content). The constraint (\ref{first:h})  ensures that converting $i$ to $i'$ is only executed at  one location, either  in self-driving car $v$ or at MEC server $r$.

The formulated optimization problem in (\ref{eq:problem_formulation}) is non-convex problem which makes it complicated to solve. Therefore, in the next Subsection \ref{subsec:ProposedSolution}, we propose a proximal convex surrogate problem of the formulated problem in (\ref{eq:problem_formulation}) and apply Block Successive Majorization-Minimization (BS-MM) \cite{sun2017majorization} for solving proximal convex surrogate problem.

\subsection{Proposed Solution: Distributed Algorithm for Deep Learning Based Caching}
\label{subsec:ProposedSolution} 

For solving our optimization problem, we use BS-MM described in \cite{sun2017majorization, ndikumana2018deep}.  We chose BS-MM over other distributed algorithms such as DC (Difference of Convex) programming, concave-convex, and successive convex approximation because BS-MM is a new approach that allows to partition the problem into blocks and applies MM to one block of variables while keeping the values of the other blocks fixed \cite{sun2017majorization}. The BS-MM may have computation overhead due to the computation of the best solution at each iteration, especially when the size of the problem is very large. Also, when BS-MM is fast, it may skip the true local minimum. If BS-MM is too slow, it may never converge because it tries to find a local minimum at each iteration. 
Therefore, to overcome these BS-MM challenges and ensure that all blocks are utilized, as suggested in \cite{hong2017iteration},  we use different selection rules such as Cyclic, Gauss-Southwell, and Randomized described in \cite{hong2017iteration}. To apply BS-MM in (\ref{eq:problem_formulation}),  we consider $\mathcal{Q}\triangleq\{\vect{q}:\sum_{u=1}^Uq^v_u \leq 1,\; q^v_u \in [0,1]\} $,  $\mathcal{H}\triangleq\{\vect{h}:\sum_{u=1}^U(h_i^{u\rightarrow v} + h_{i'}^{v\rightarrow u})+ \left(1-(h_i^{u\rightarrow v} + h_{i'}^{v\rightarrow u})\right)(h_{i}^{r\rightarrow v}+h_{i'}^{r\rightarrow v}) \leq  1, h_i^{u\rightarrow v}, h_{i'}^{v\rightarrow u},h_{i}^{r\rightarrow v}, h_{i'}^{r\rightarrow v}  \in [ 0, 1]\}$, and $ \mathcal{P}\triangleq \{\vect{\varrho}: \sum_{i,i' \in \mathcal{I}} \varrho_{v}^{i\rightarrow{i'}}  + (1-\varrho_{v}^{i\rightarrow{i'}})\varrho_{r}^{i\rightarrow{i'}}\leq 1 , \varrho_{v}^{i\rightarrow{i'}}, \varrho_{r}^{i\rightarrow{i'}}  \in [0,1]\}$ as non-empty and closed sets of the relaxed variables $\vect{q}$,  $\vect{h}$, and $\vect{\varrho}$, respectively. Therefore, 
to simplify our notation, we use $\mathcal{F}(\vect{q}, \vect{h}, \vect{\varrho})$ to denote (\ref{eq:problem_formulation}), where $\mathcal{F}(\vect{q}, \vect{h}, \vect{\varrho})$ is expressed as follows:

\begin{equation}
\label{eq:new_problem_formulation}
\mathcal{F}(\vect{q}, \vect{h}, \vect{\varrho})=\sum_{u=1}^U\tau^\textrm{Tot}_{u}(\vect{q}, \vect{h}, \vect{\varrho}). 
\end{equation}

Both  (\ref{eq:problem_formulation}) and (\ref{eq:new_problem_formulation}) have the same constraints. Therefore, to solve (\ref{eq:new_problem_formulation}), we use the following steps:
\begin{itemize}
	\item
	In the first step, called majorization, we propose  a proximal convex surrogate problem $\mathcal{F}_j(\vect{q}, \vect{h}, \vect{\varrho})$ (\ref{eq:proximal_problem}) of the formulated problem in (\ref{eq:new_problem_formulation}), which is an upper-bound of (\ref{eq:new_problem_formulation}).  
	\item
	In the second step, called minimization, instead of minimizing (\ref{eq:new_problem_formulation}) which is intractable, we minimize its proximal convex surrogate function $\mathcal{F}_j(\vect{q}, \vect{h}, \vect{\varrho})$(\ref{eq:proximal_problem}).        
\end{itemize}

The success  of BS-MM relies on the surrogate function. Therefore, a surrogate function that is easy to solve and upper-bound of of the formulated problem in (\ref{eq:new_problem_formulation}) is preferable. To achieve this, in the majorization step, we use the proximal upper-bound minimization technique described in \cite{sun2017majorization}. Then, we propose the following proximal convex surrogate problem $\mathcal{F}_j(\vect{q}, \vect{h}, \vect{\varrho})$ (\ref{eq:proximal_problem}) of the formulated problem in (\ref{eq:new_problem_formulation}) by adding the quadratic term ($\frac{ \varrho_j}{2} \lVert(\vect{q}_j- \vect{q}^{(0)})\rVert^2$) to (\ref{eq:new_problem_formulation}):
\begin{equation}
\label{eq:proximal_problem}
\mathcal{F}_j(\vect{q}_j, \vect{q}^{(t)},\vect{h}^{(t)}, \vect{\varrho}^{(t)})\defeq \mathcal{F}{(\vect{q}_j,\vect{q}^{(0)}, \vect{h}^{(0)}, \vect{\varrho}^{(0)}}) + \frac{ \alpha_j}{2} \lVert(\vect{q}_j- \vect{q}^{(0)})\rVert^2, 
\end{equation}
where $\vect{q}^{(0)}$, $\vect{h}^{(0)}$, and $\vect{\varrho}^{(0)}$  are the initial feasible points.  Furthermore, the surrogate function in (\ref{eq:proximal_problem}) can be applied to other vectors  $\vect{h}$ and $\vect{\varrho}$. In addition, the quadratic term ($\frac{ \alpha_j}{2} \lVert(\vect{q}_j- \vect{q}^{(0)})\rVert^2$) makes the problem (\ref{eq:proximal_problem}) to be convex and upper-bound of (\ref{eq:new_problem_formulation}). In the minimization step, we minimize the surrogate function $\mathcal{F}_j(\vect{q}, \vect{h}, \vect{\varrho})$ (\ref{eq:proximal_problem}) by taking steps proportional to the negative of the gradient in the direction toward the formulated problem in (\ref{eq:new_problem_formulation}), where $\mathcal{J}^t$ is a set of indexes at each iteration $t$ and $\alpha_j$ is a positive penalty parameter for $j \in \mathcal{J}^t$. At each iteration $t + 1$, the solution is updated by solving the following problems: 
\begin{equation}
\begin{aligned}
\vect{q}_j^{(t+1)}\in \underset{ \vect{q}_j \in \mathcal{Q}}{\text{min}}\;\mathcal{F}_j(\vect{q}_j, \vect{q}^{(t)},\vect{h}^{(t)}, \vect{\varrho}^{(t)}),
\end{aligned}
\label{eq:optimization20}
\end{equation} 
\begin{equation}
\begin{aligned}
\vect{h}_j^{(t+1)}\in \underset{ \vect{h}_j \in \mathcal{H}}{\text{min}}\; \mathcal{F}_j(\vect{h}_j, \vect{h}^{(t)},\vect{q}_j^{(t+1)}, \vect{\varrho}^{(t)}),
\end{aligned}
\label{eq:optimization21}
\end{equation}
\begin{equation}
\begin{aligned}
\vect{\varrho}_j^{(t+1)}\in \underset{ \vect{\varrho}_j \in \mathcal{P}}{\text{min}}\; \mathcal{F}_j(\vect{\varrho}_j, \vect{\varrho}^{(t)},\vect{q}_j^{(t+1)}, \vect{h}_j^{(t+1)}).
\end{aligned}
\label{eq:optimization22}
\end{equation} 

To solve our problems in (\ref{eq:optimization20}), (\ref{eq:optimization21}), and (\ref{eq:optimization22}) we use vectors  $\vect{q}_j$, $\vect{h}_j$ and $\vect{\varrho}_j$ of relaxed variables. Therefore, we need to enforce $\vect{q}_j$, $\vect{h}_j$ and $\vect{\varrho}_j$ to be vectors of  binary variables.  To achieve this, we apply the rounding techniques described in \cite{feige2016oblivious}. As an illustration example, for  a solution $ q^{r*}_v \in \vect{q}_j^{(t+1)}$,  we define the rounding threshold $\varphi \in (0,1)$, such that the enforced binary value of $ q^{r*}_v$ is given by:  
\begin{equation}
\label{eq:b_rounding}
\setlength{\jot}{10pt}
q^{r*}_v =
\begin{cases}
1,\; \text{if $q^{r*}_v \geq \varphi $},\\
0, \;\text{otherwise.}
\end{cases}
\end{equation}

As highlighted in \cite{ndikumana2018joint, zhang2017network}, the rounding technique may violate 3C resource constraints. Therefore, to overcome this issue,  we solve $\mathcal{F}_j$ in the form $\mathcal{F}_j+\beta_v\Delta_v$ by updating the constrains in (\ref{first:d}), (\ref{first:e}), and (\ref{first:f}) as follows:
\begin{equation}
\sum_{v=1}^Vq^r_v a_v^r \leq 1 + \Delta_{v_a}, \;  \forall r \in \mathcal{R},\label{first:a_m}
\end{equation}    
\begin{equation}
\label{first:b_m}
\sum_{u=1}^U\sum_{i=1}^{I_r(n)}q^v_u h_i^{u\rightarrow v}\varrho_{v}^{i\rightarrow{i'}} p^{i\rightarrow {i'}}_v \leq p_v+ \Delta_{v_p},\forall  v \in \mathcal{V},
\end{equation}    
\begin{equation}
\label{first:c_m}
q^r_v\sum_{j=1}^k(\sum_{\tilde{i}_f \in \mathcal{G}_{je}^\textrm{female}}o_v^{\tilde{i}_f}S(\tilde{i}_f))  + \sum_{\tilde{i}_m \in \mathcal{G}_{je}^\textrm{male}}o_v^{\tilde{i}_m}S(\tilde{i}_m)) \leq c_v + \Delta_{v_c},
\end{equation}
where $\Delta_v=\Delta_{v_a}+\Delta_{v_p}+\Delta_{v_c}$ is the maximum violation of the 3C resource constraints and $\beta_v$ as the weight parameter of $\Delta_v$. Furthermore, the values of $\Delta_{v_a}$, $\Delta_{v_p}$, and $\Delta_{v_c}$ are given by:
\begin{equation}
\label{eq:rounding-10}
\Delta_{v_a} =\max \{ 0,\sum_{v=1}^Vq^r_v a_v^r - 1 \},\;\forall r \in \mathcal{R},
\end{equation}
\begin{equation}
\label{eq:rounding-11}
\Delta_{v_p} = \max \{0, \sum_{u=1}^U\sum_{i=1}^{I_r(n)}q^v_u h_i^{u\rightarrow v}\varrho_{v}^{i\rightarrow{i'}} p^{i\rightarrow {i'}}_v  - p_v  \},\;\forall  v \in \mathcal{V},
\end{equation}
\begin{multline}
\label{eq:rounding-22}
\Delta_{v_c}=\max \{0, q^r_v\sum_{j=1}^k((\sum_{\tilde{i}_f \in \mathcal{G}_{je}^\textrm{female}}o_v^{\tilde{i}_f}S(\tilde{i}_f))  + \\\sum_{\tilde{i}_m \in \mathcal{G}_{je}^\textrm{male}}o_v^{\tilde{i}_m}S(\tilde{i}_m))- c_v  \}.
\end{multline}

Therefore, to ensure that the best solution is achieved, we use the integrality gap described in \cite{feige2016oblivious}. 

\begin{definition}[Integrality gap] For the problems $\mathcal{F}_j+\beta_v\Delta_v$ and $\mathcal{F}_j$, the integrality gap is expressed as follows:
	\begin{equation}
	\phi_j=\underset{\vect{q}, \vect{h}, \vect{\varrho}}{\text{min}}\ \ \frac{\mathcal{F}_j}{\mathcal{F}_j+\beta_v\Delta_v}.
	\end{equation}
\end{definition}
The best solutions of $\mathcal{F}_j$ and $\mathcal{F}_j+\beta_v\Delta_v$ are obtained when $\phi_j \leq 1$.

\begin{algorithm}[t]    
	\caption{: Distributed algorithm for deep learning based caching.}
	\label{algo:onlienalgorithm}
	\begin{algorithmic}[1]
		\STATE{\textbf{Preconditions:} MLP output and CNN models are deployed to the RSUs and in self-driving car};
		\STATE{\textbf{Input:} $\vect{U}$: A vector of passengers, $\vect{\omega}^r_{v}$:   wireless link capacities, $\vect{\mathcal{X}}$: Vector of recommended contents for $\mathcal{G}_{je}^\textrm{female}$ and $\mathcal{G}_{je}^\textrm{male}$ in self-driving car $v$,  $\psi^v_u$, $p_v$, and $c_v$; }
		\STATE{\textbf{Output:} $\vect{q}^*, \;\vect{h}^*, \;\vect{\varrho}^*$};
		\STATE {Initialize $t=0$;}
		\STATE {Find initial feasible points $\vect{q}^{(0)}$,  $\vect{h}^{(0)}$, $\vect{\varrho}^{(0)}$;}
		\REPEAT
		\STATE{Choose index set $\mathcal{J}^t$;}
		\STATE{Let $\vect{q}_j^{(t+1)}\in \underset{ \vect{q}_j \in \mathcal{Q}}{\text{min}}\;\mathcal{F}_j(\vect{q}_j, \vect{q}^{(t)},\vect{h}^{(t)}, \vect{\varrho}^{(t)})$ (\ref{eq:optimization20});}
		\STATE{Set $\vect{q}_k^{t+1}=\vect{q}_k^{t}, \forall k \notin  \mathcal{J}^t$ and solve $\underset{ \vect{q}_j \in \mathcal{Q}}{\text{min}}\;\mathcal{F}_j(\vect{q}_j, \vect{q}^{(t)},\vect{h}^{(t)}, \vect{\varrho}^{(t)})$;}
		\STATE{ For $\vect{h}_j^{(t+1)}$ and $\vect{\varrho}_j^{(t+1)}$, restart from step $4$, salve (\ref{eq:optimization21}) and (\ref{eq:optimization22});}    
		\STATE{$t=t+1$;}    
		\UNTIL{ $\lim\limits_{t \to \infty}\underset{\vect{q}, \vect{h}, \vect{\varrho}}{\text{inf}}\lVert\mathcal{F}_j^{(t+1)}-\mathcal{F}_j^{(t)}\rVert_2=0$;}
		\STATE{By rounding technique, enforce  $\vect{q}_j^{(t+1)}$, $\vect{h}_j^{(t+1)}$ , and $\vect{\varrho}_j^{(t+1)}$ to be vectors of binary variables;}
		\STATE{Solve $\mathcal{F}_j+\beta_v\Delta_v$ and compute  $\phi_j$ until $\phi_j \leq 1$;}
		\STATE{Then, consider $\vect{q}^*=\vect{q}_j^{(t+1)}$, $\vect{h}^*=\vect{h}_j^{(t+1)}$, and $\vect{\varrho}^*=\vect{\varrho}_j^{(t+1)}$ as a solution.}
	\end{algorithmic}
\end{algorithm}
%++++++++++++++++++++++++++++++++++++++++++++++++++ 
  
We propose a distributed algorithm (Algorithm \ref{algo:onlienalgorithm}), which is based on BS-MM \cite{sun2017majorization}. We assume that the MLP output and CNN model are already deployed at RSUs and in self-driving car. We consider  a vector of passengers, vector of RSUs,  vector of wireless link capacities, vector of recommended contents that need to be cached in self-driving car $v$,  $\psi^v_u$, $p_v$, and $c_v$ as the  input. First, Algorithm \ref{algo:onlienalgorithm} finds the initial feasible points $\vect{q}^{(0)}$, $\vect{h}^{(0)}$, and $\vect{\varrho}^{(0)}$. Then, Algorithm \ref{algo:onlienalgorithm} starts an iterative process by choosing an index set $\mathcal{J}^t$ at each iteration $t$. At each iteration $t+1$, the solution is updated by solving the problems (\ref{eq:optimization20}), (\ref{eq:optimization21}), and (\ref{eq:optimization22})  until $\lim\limits_{t \to \infty}\underset{\vect{q}, \vect{h},\vect{\varrho}}{\text{inf}}\lVert\mathcal{F}_j^{(t+1)}-\mathcal{F}_j^{(t)}\rVert_2=0$, where  $\lim\limits_{t \to \infty}\underset{\vect{q}, \vect{h},\vect{\varrho}}{\text{inf}}\lVert\mathcal{F}_j^{(t+1)}-\mathcal{F}_j^{(t)}\rVert_2=0$ is the convergence criteria.  Therefore, when $\lim\limits_{t \to \infty}\underset{\vect{q}, \vect{h},\vect{\varrho}}{\text{inf}}\lVert\mathcal{F}_j^{(t+1)}-\mathcal{F}_j^{(t)}\rVert_2=0$, Algorithm \ref{algo:onlienalgorithm} considers $\vect{q}_j^{(t+1)}$, $\vect{h}_j^{(t+1)}$, and $\vect{\varrho}_j^{(t+1)}$ as a solution. Then, Algorithm \ref{algo:onlienalgorithm} forces the solution $\vect{q}_j^{(t+1)}$, $\vect{h}_j^{(t+1)}$, and $\vect{\varrho}_j^{(t+1)}$ to be vectors of binary variables via the rounding technique, where Algorithm \ref{algo:onlienalgorithm} solves $\mathcal{F}_j+\beta_v\Delta_v$ and computes $\phi_j$. Finally, when  $\phi_j \leq 1$, Algorithm \ref{algo:onlienalgorithm}  considers $\vect{q}^*=\vect{q}_j^{(t+1)}$, $\vect{h}^*=\vect{h}_j^{(t+1)}$, and $\vect{\varrho}^*=\vect{\varrho}_j^{(t+1)}$ as a solution which does not violate 3C resource constraints. Furthermore, for the convergence of the proposed algorithm, based on the convergence of MM defined and proved in \cite{sun2017majorization}, we make the following remark:

\begin{remark}[Convergence of the proposed algorithm] Based on the MM algorithm \cite{sun2017majorization}, the proposed Algorithm \ref{algo:onlienalgorithm}, which is based on BS-MM, converges to coordinate-wise minimum point which is stationary point, when the vectors  $\vect{q}^*=\vect{q}_j^{(t+1)}$, $\vect{h}^*=\vect{h}_j^{(t+1)}$, and $\vect{\varrho}^*=\vect{\varrho}_j^{(t+1)}$ cannot find a better minimum direction, i.e., $\lim\limits_{t \to \infty}\underset{\vect{q}, \vect{h}, \vect{\varrho}}{\text{inf}}\lVert\mathcal{F}_j^{(t+1)}-\mathcal{F}_j^{(t)}\rVert_2=0$.
\end{remark} 
\begin{table}[!t]
	\centering
	\caption{The used route for the self-driving bus.}
	\label{tab:Evaluationsetting1}
	\begin{tabular}[t]{ |l|l|l|l|l|l|}
		\hline
		Route &Distance (Km) &Max. speed (Km/h)& RSUs\\
		\hline
		$1$ & $54.62$  & $109.016$ & $1-2$ \\ 
		\hline
		$2$ & $53.82$ &  $107.34$ & $2-3$  \\ 
		\hline
		$3$ & $54.02$ & $108.17$ & $3-4$ \\ 
		\hline
		$4$ & $52.83$ &  $105.38$ & $4-5$  \\ 
		\hline
		$5$ & $55.66$  & $111.33$ & $5-6$ \\ 
		\hline	
	\end{tabular}
\end{table}%

For complexity analysis of the proposed Algorithm \ref{algo:onlienalgorithm}, based on complexity analysis described in \cite{hong2017iteration}, we make the following remark:

\begin{remark}[ Complexity of the proposed Algorithm \ref{algo:onlienalgorithm}] The Algorithm \ref{algo:onlienalgorithm}, which is based on BS-MM, uses proximal upper-bound minimization technique. This makes it fall under the BSUM framework \cite{hong2017iteration}. Therefore, for the iteration index $j \in \mathcal{J}^t$,  the Algorithm \ref{algo:onlienalgorithm} has $\mathcal{O}(1/j)$ iteration complexity, which is sub-linear.
\end{remark} 

\section{Simulation Results and Analysis}
\label{sec:PerformanceEvaluation}

In this section, we present a performance evaluation of the proposed deep learning-based caching in self-driving cars. We use Google Maps Services \cite{googlemaps} for the self-driving car mobility analysis, Keras with Tensorflow \cite{Keras} for the deep learning simulation, and pandas \cite{mckinney2011pandas} for data analysis.

\subsection{Simulation Setup}
\label{subsec:PerformanceEvaluation}

\begin{figure}[t]
	\begin{minipage}{0.45\textwidth}		
		\centering
		\includegraphics[width=0.95\columnwidth]{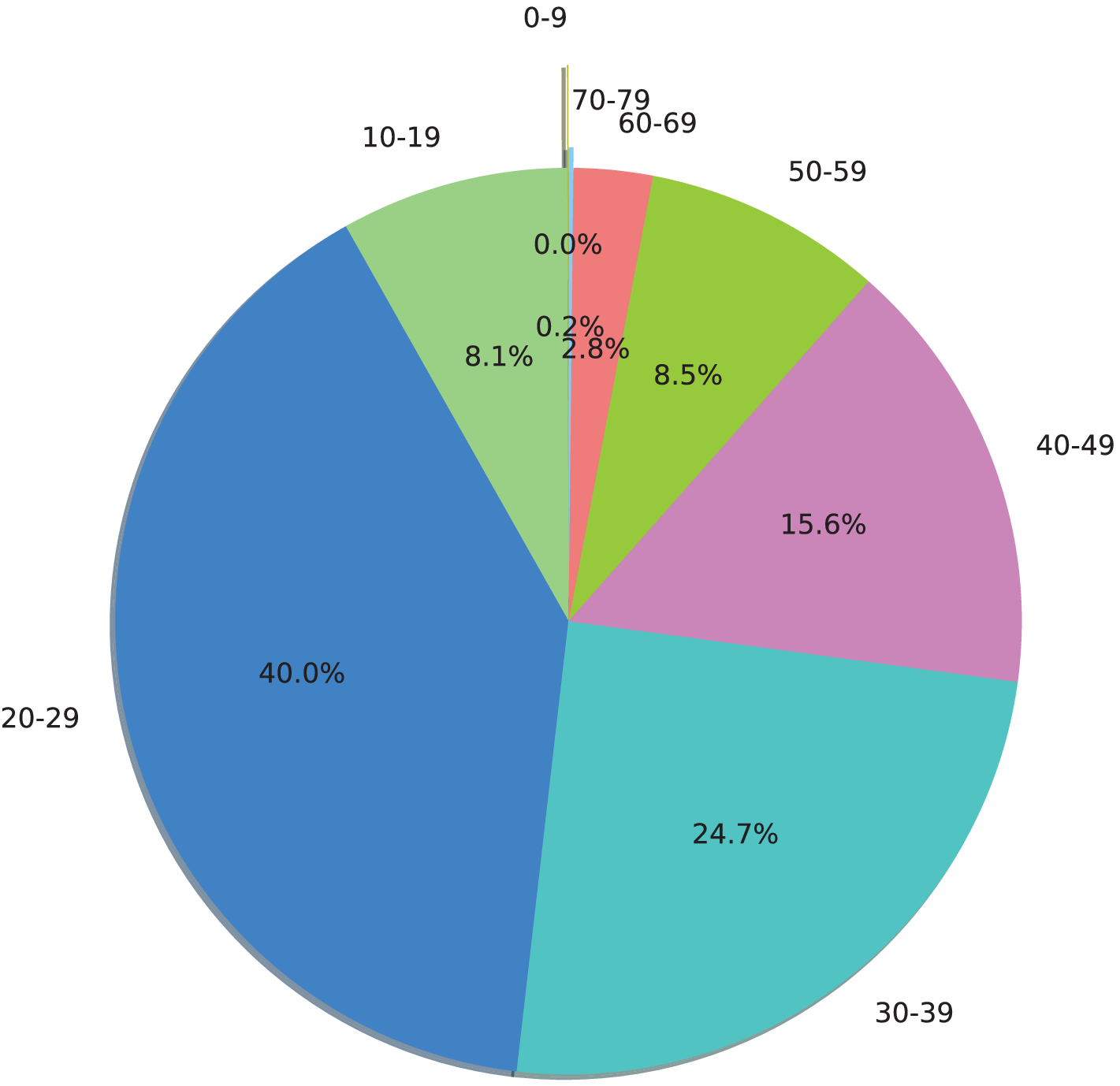}
		\caption{Visualization of the used dataset \cite{harper2016movielens} for movie watching based on age.}
		\label{fig:video_watch_count}
	\end{minipage}
	\begin{minipage}{0.45\textwidth}		
		\centering
		\includegraphics[width=1.00\columnwidth]{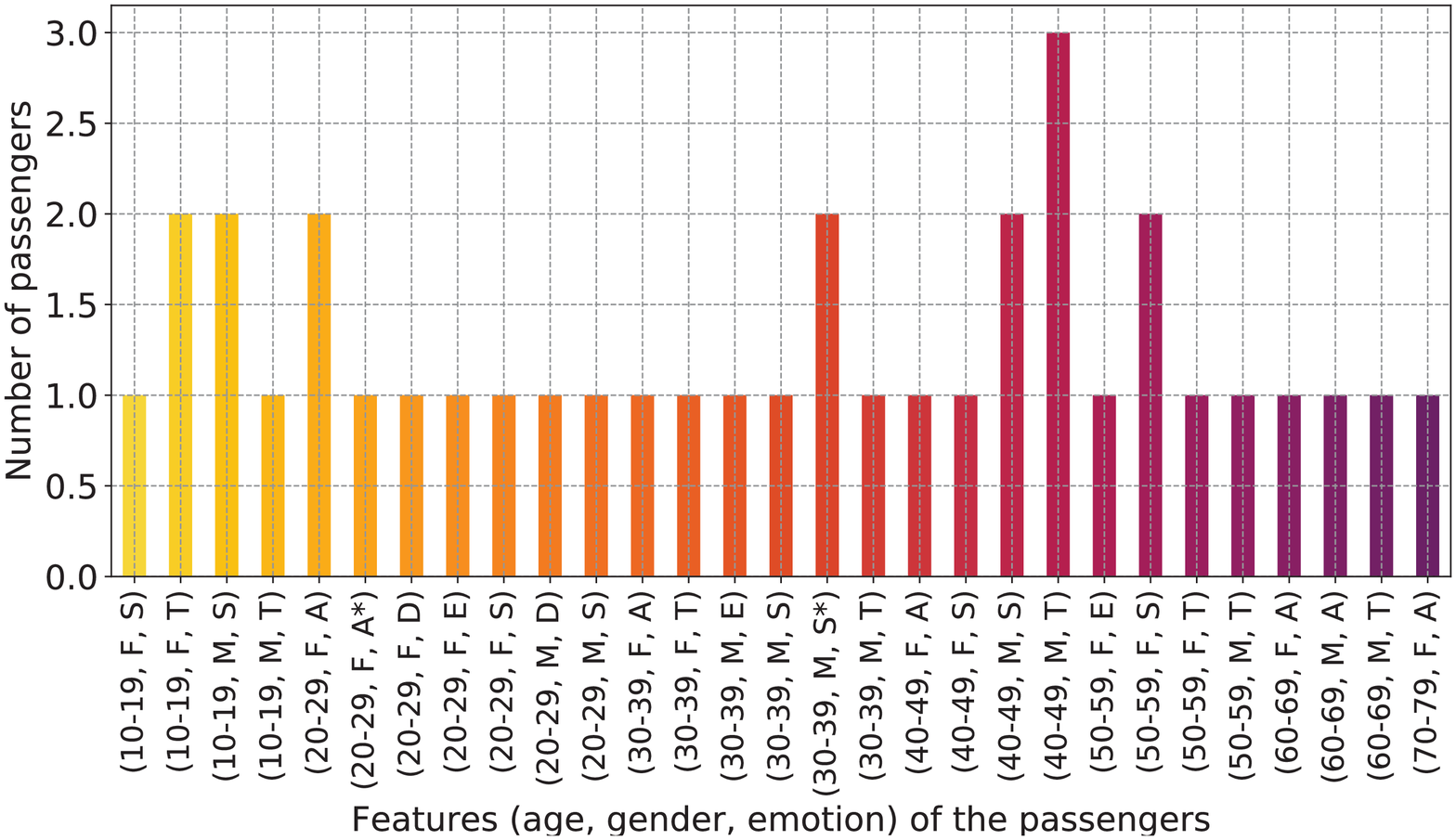}
	\caption{Visualization of the used  passengers' features for the self-driving bus.}
		\label{fig:self_driving_passenger}
	\end{minipage}
\end{figure}

To predict the probabilities of contents to be requested in specific areas of MEC servers, we use a  well-known dataset called Movie-Lens Dataset \cite{harper2016movielens}. \textcolor{black}{In the dataset,  we have movies with related information such as movie titles, release date, and genre of movies such as comedy, drama, and documentary. We associate the emotion with the genre of movies, where sad  users recommended to watch drama movies, disgust users recommended to watch musical movies, anger users recommended to watch comedy movies, anticipate users recommended to watch thriller movies, fear users recommended to watch adventure movies, joy users recommended to watch thriller movies, trust users recommended to watch western movies, and surprise users recommended to watch  fantasy movies.}  However, the dataset does not have movie sizes and formats. Since our deep learning-based caching scheme uses content size, we randomly generate size $S(i)$ for each movie $i$ in the range from $S(i)=317$ to $S(i)=750$ Mb and randomly assign each movie $i$ a format. Furthermore, we have user's information such as age (as shown in Fig. \ref{fig:video_watch_count}), gender,  rating, and ZIP codes. To identify the areas of users, we convert the ZIP codes into longitude and latitude coordinates and deploy $6$ RSUs to the specific areas based on the movie watching counts, rankings, and the locations of users. We use MLP  with $2$ layers (for input and output) and $2$ hidden layers to predict the probabilities of contents to be requested in specific areas of RSUs. In MLP,  each layer has $32$ neurons except the output layer which has $6$ neurons.  In the output layer, $6$ neurons correspond to the probabilities of contents to be cached in specific areas of $6$ RSUs. We use $60\%$ of the dataset for training and $40\%$ for testing. Furthermore, the learning rate is set to be equal to $0.002$, while the batch size equals to $32$.
\begin{figure}[t]
	\begin{minipage}{0.45\textwidth}
		\centering
		\includegraphics[width=0.95\columnwidth]{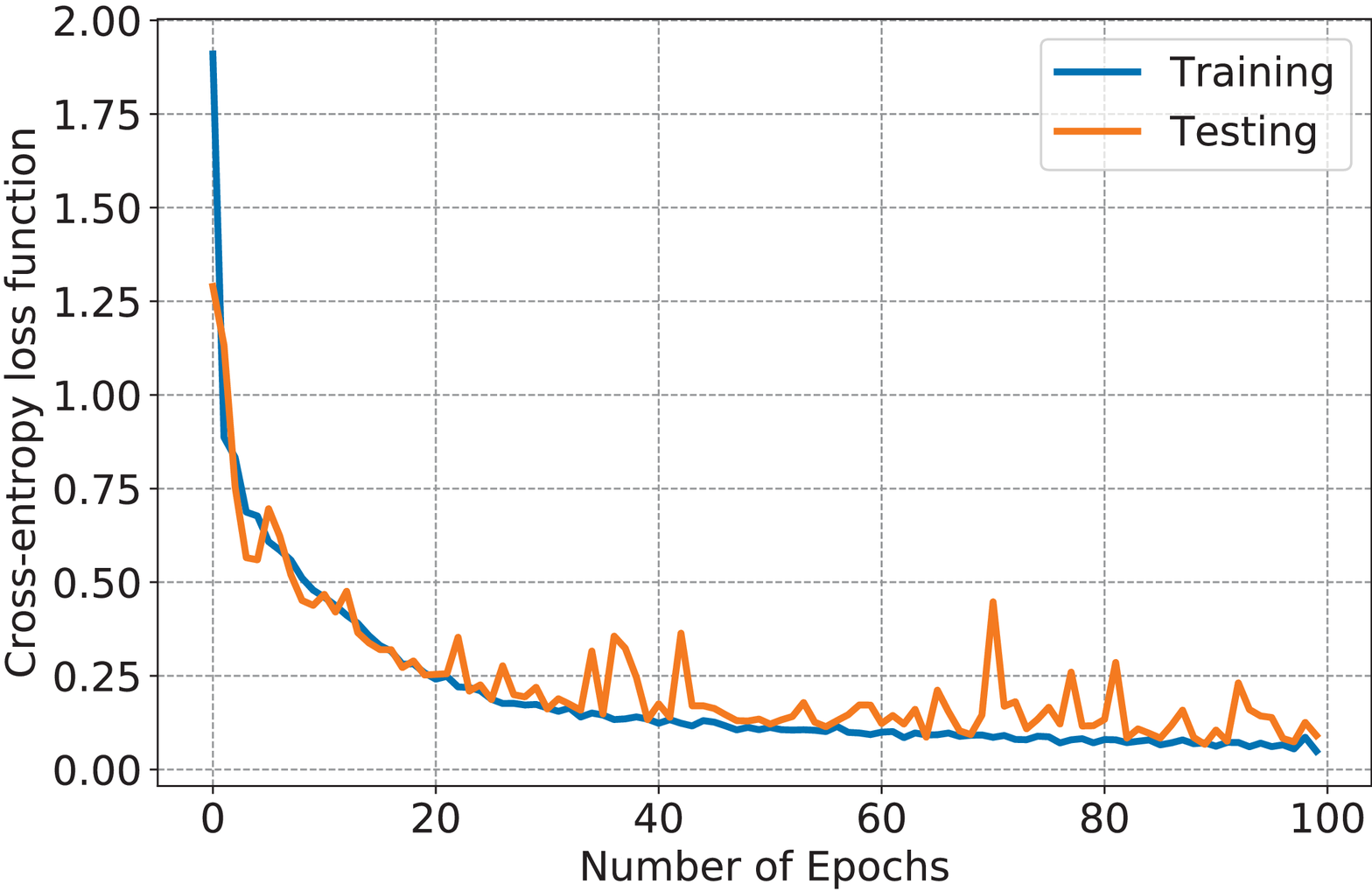}
		\caption{Minimization of error function for predicting the probability of movies to be requested in the specific areas of RSUs (acc: $97.82\%$).}
		\label{fig:video_request_probabilities}
	\end{minipage}
	\begin{minipage}{0.45\textwidth}	
	\centering
	\includegraphics[width=0.95\columnwidth]{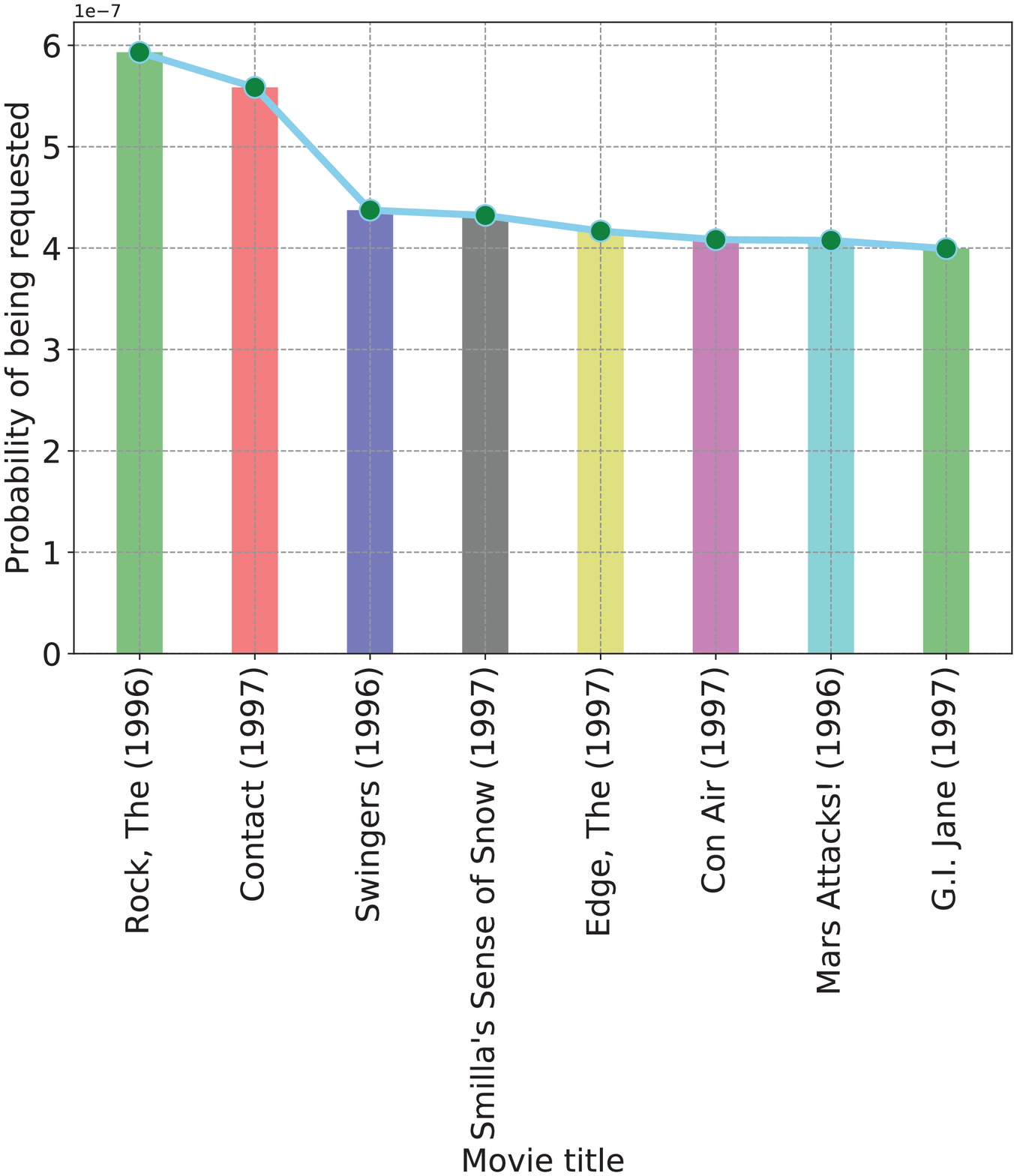}
	\caption{ Some high recommended movies to cache in close proximity of the self-driving cars at RSUs.}
	\label{fig:cache_probabilities}.
\end{minipage}

\end{figure}

With the departure time and locations of the RSUs, the Google Maps service provides the distance and duration to reach each RSU $r\in \mathcal{R}$, where the duration is based on traffic conditions between the source and destination. Based on the distance (in terms of km) and duration (in terms of hours), we can calculate the speed (in terms of km/h) of the self-driving car and find the RSUs that the self-driving car can connect to for retrieving contents. However, based on Google Maps service \cite{googlemaps}, the distances between RSUs are very large. Therefore, to have realistic distances between RSUs,  we update the RSU locations and create a routing table summarized in Table \ref{tab:Evaluationsetting1}, where the self-driving car starts its journey at RSU $1$ and ends at RSU $6$.  We set each RSU $r\in \mathcal{R}$ to be connected to the DC with a wired backhaul of capacity ranging from $\omega_{r, DC}=60$ to $\omega_{r, DC}=70$  Mbps. We assume that each RSU $r\in \mathcal{R}$ has a bandwidth of $\omega_{v, r}=10$ MHz. On the other hand, each MEC server $r\in \mathcal{R}$ has a CPU of capacity $p_r=3.6$ GHz, while the cache capacity ranges from $c_r=100$ to $c_r=110$ terabytes (TB). 

\begin{figure}[t]
	\begin{minipage}{0.45\textwidth}		
		\centering
		\includegraphics[width=1.00\columnwidth]{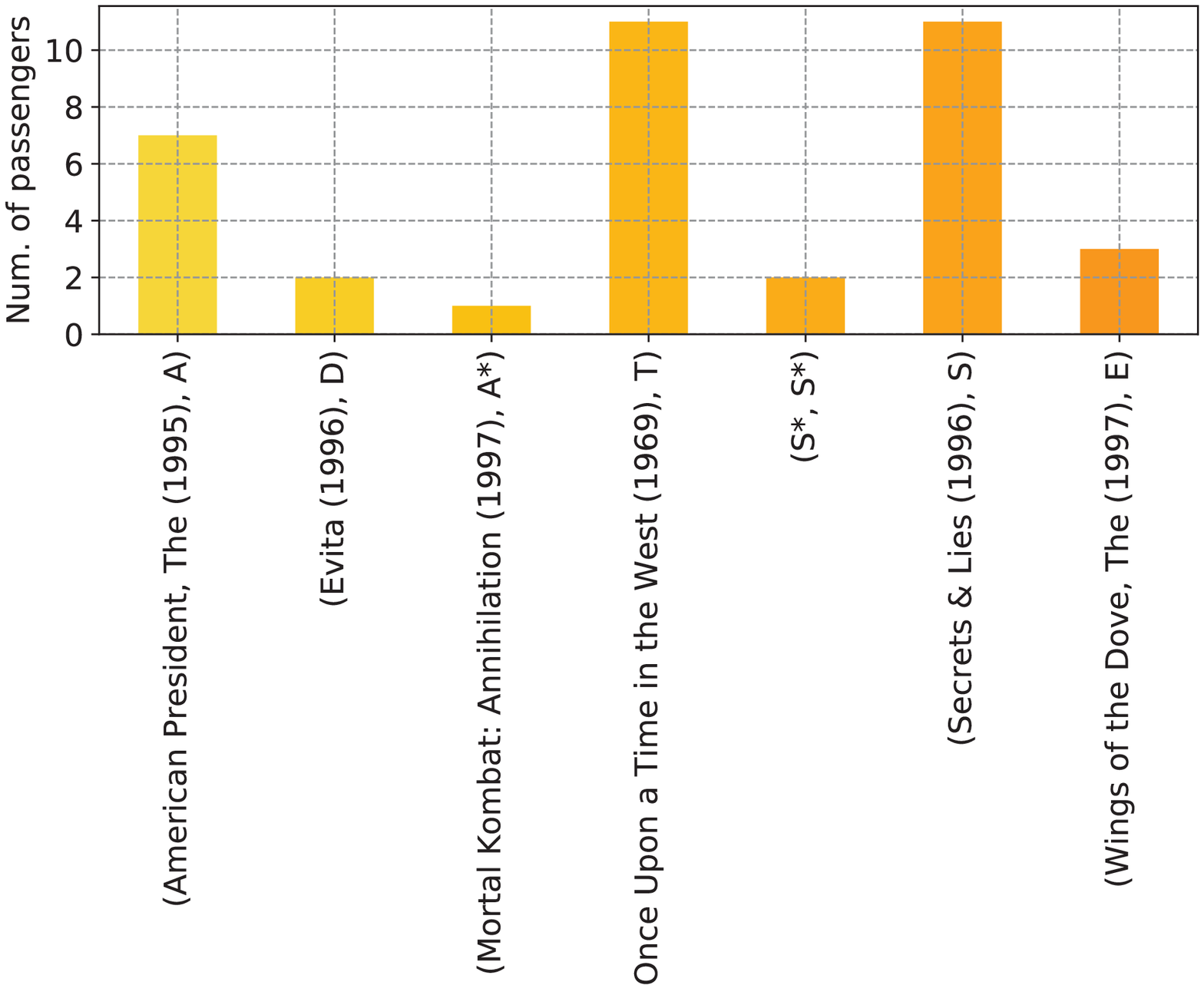}
		\caption{\textcolor{black}{Some high recommended movies to watch based on passengers' features (age, gender, and emotion).}}
		\label{fig:car_cache_recommendation}
	\end{minipage}
\begin{minipage}{0.45\textwidth}
	\centering
	\includegraphics[width=1.0\columnwidth]{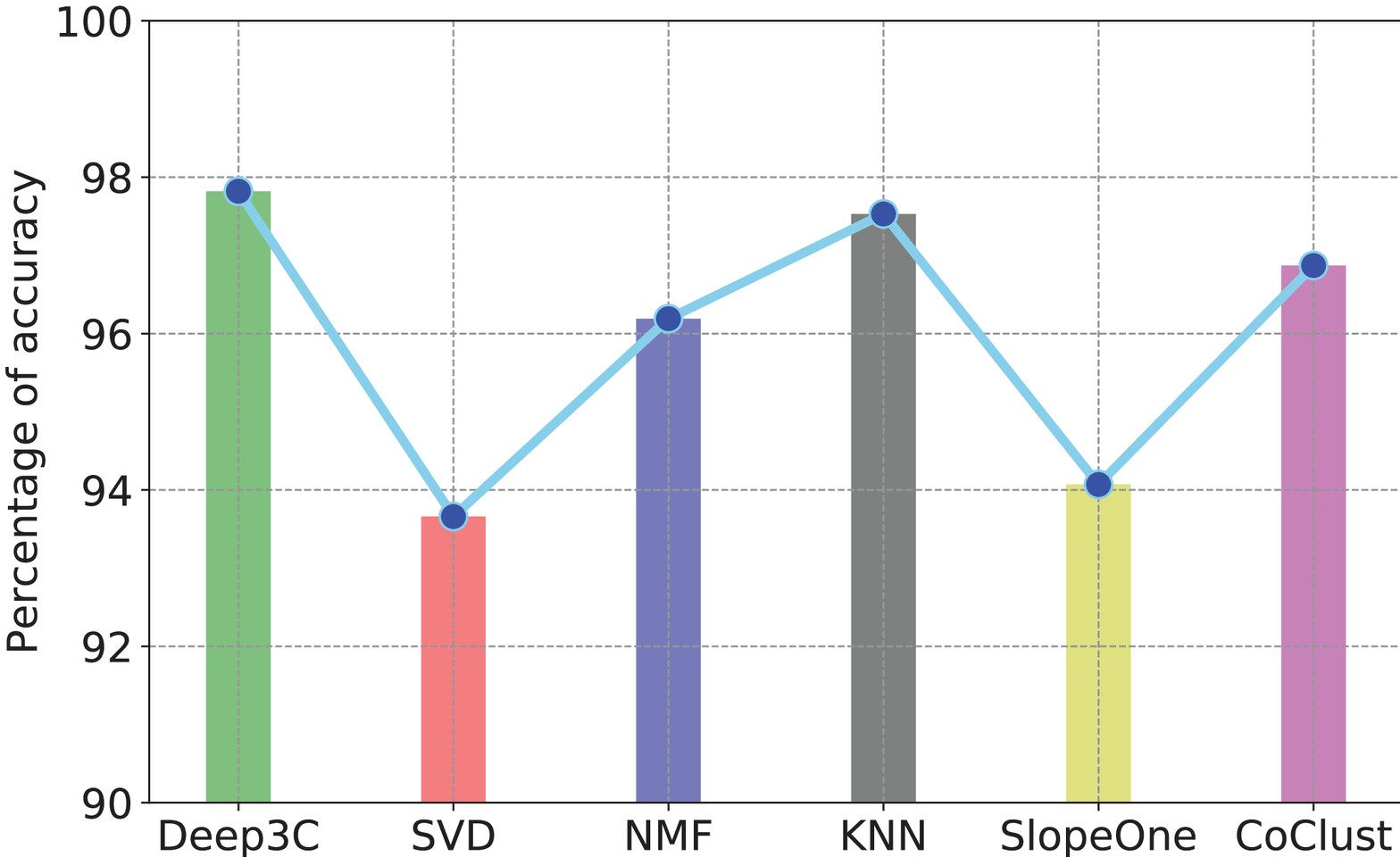}
	\caption{Comparison of various collaborative filtering algorithms and our proposal (Deep3C).}
	\label{Algorithm_comparison}
\end{minipage}
\end{figure}

\begin{figure}[t]
\begin{minipage}{0.45\textwidth}	
	\centering
	\includegraphics[width=0.95\columnwidth]{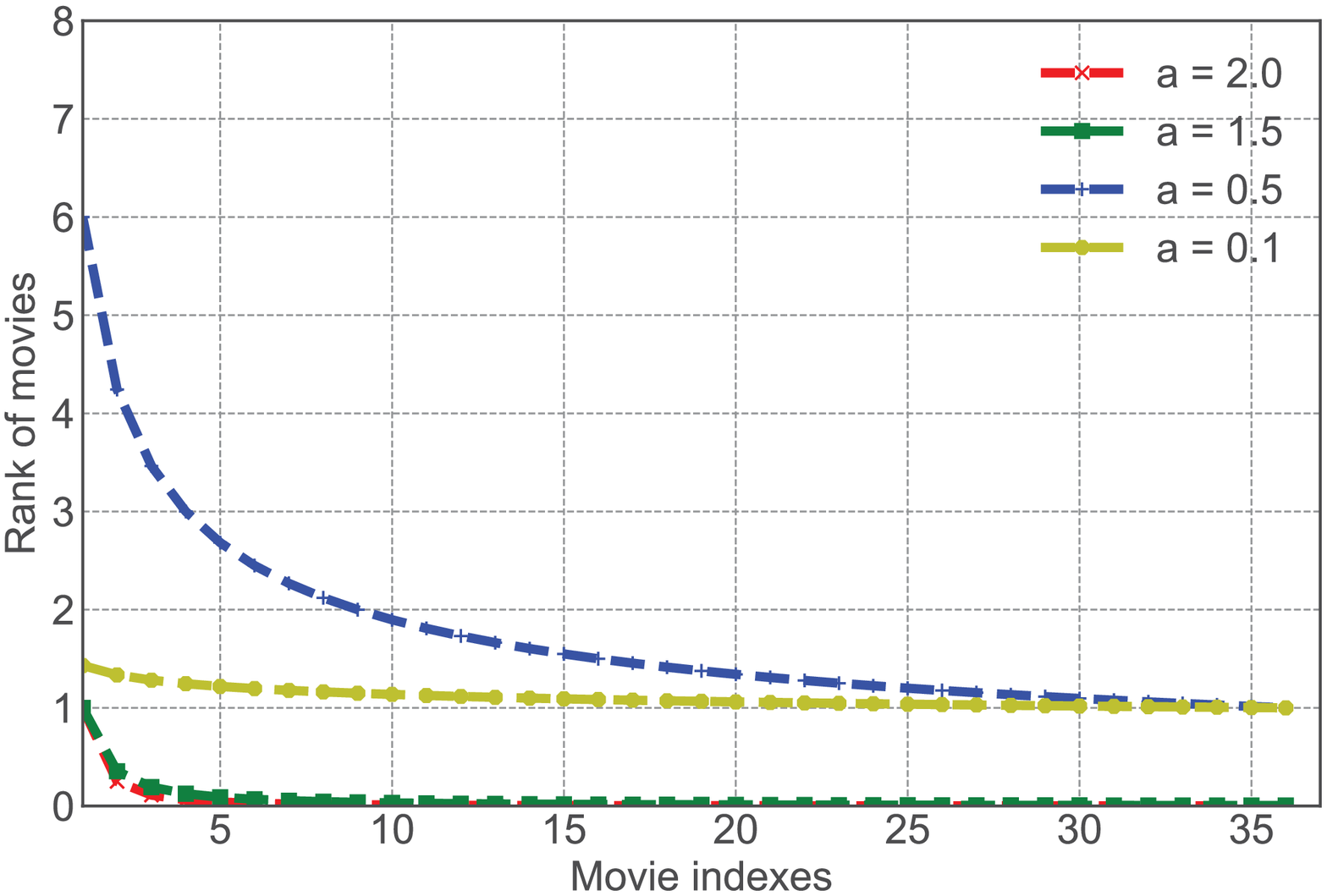}
	\caption{Ranking of movie demands based on Zipf distribution.}
	\label{fig:content_distribution}
\end{minipage}
	\begin{minipage}{0.45\textwidth}	
	\centering
	\includegraphics[width=0.95\columnwidth]{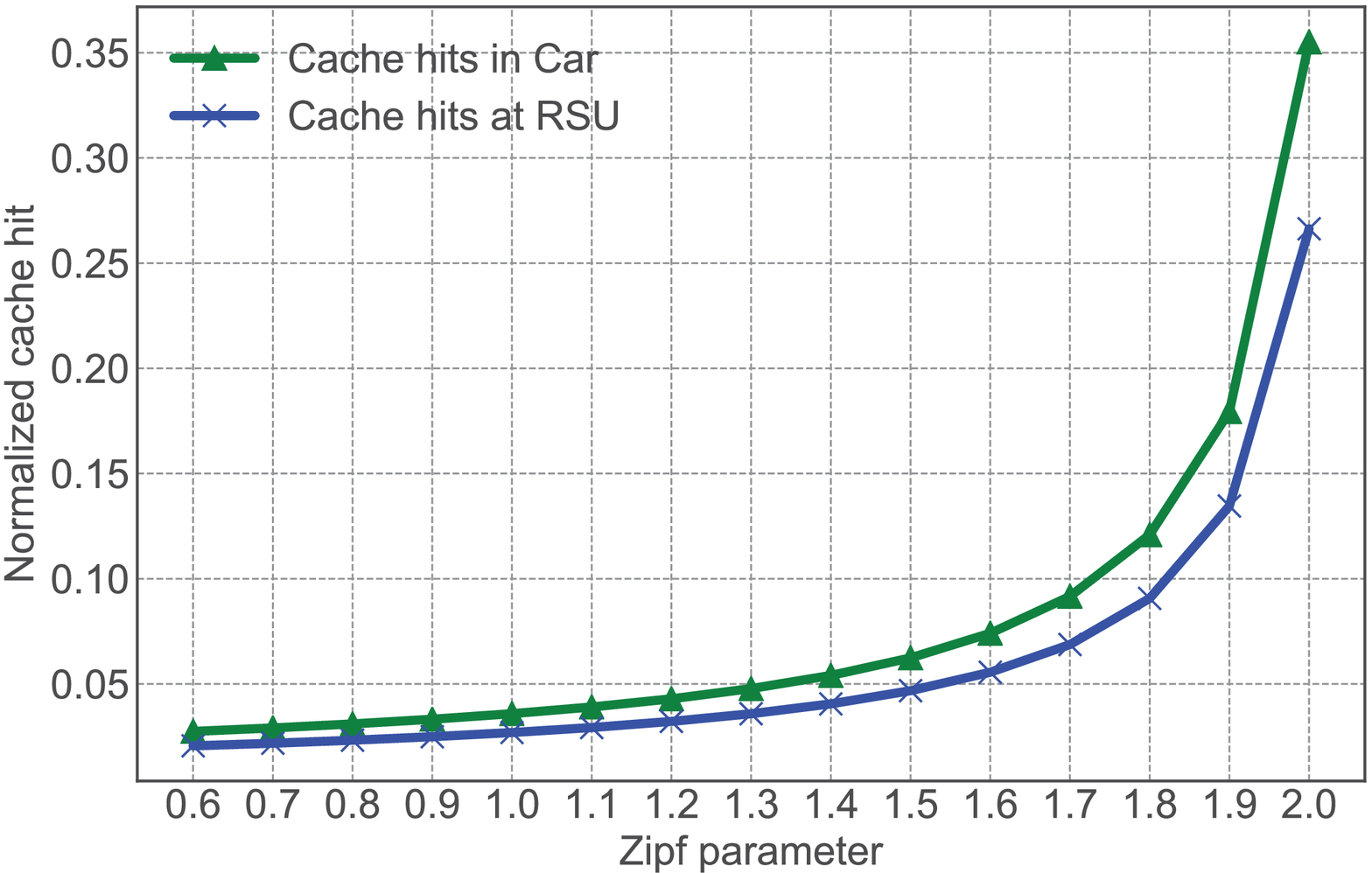}
	\caption{Cache hits for the requested movies.}
	\label{fig:cache_hit_car}
\end{minipage}	
\end{figure}

\begin{figure}[t]
	\begin{minipage}{0.45\textwidth}	
		\centering
		\includegraphics[width=0.95\columnwidth]{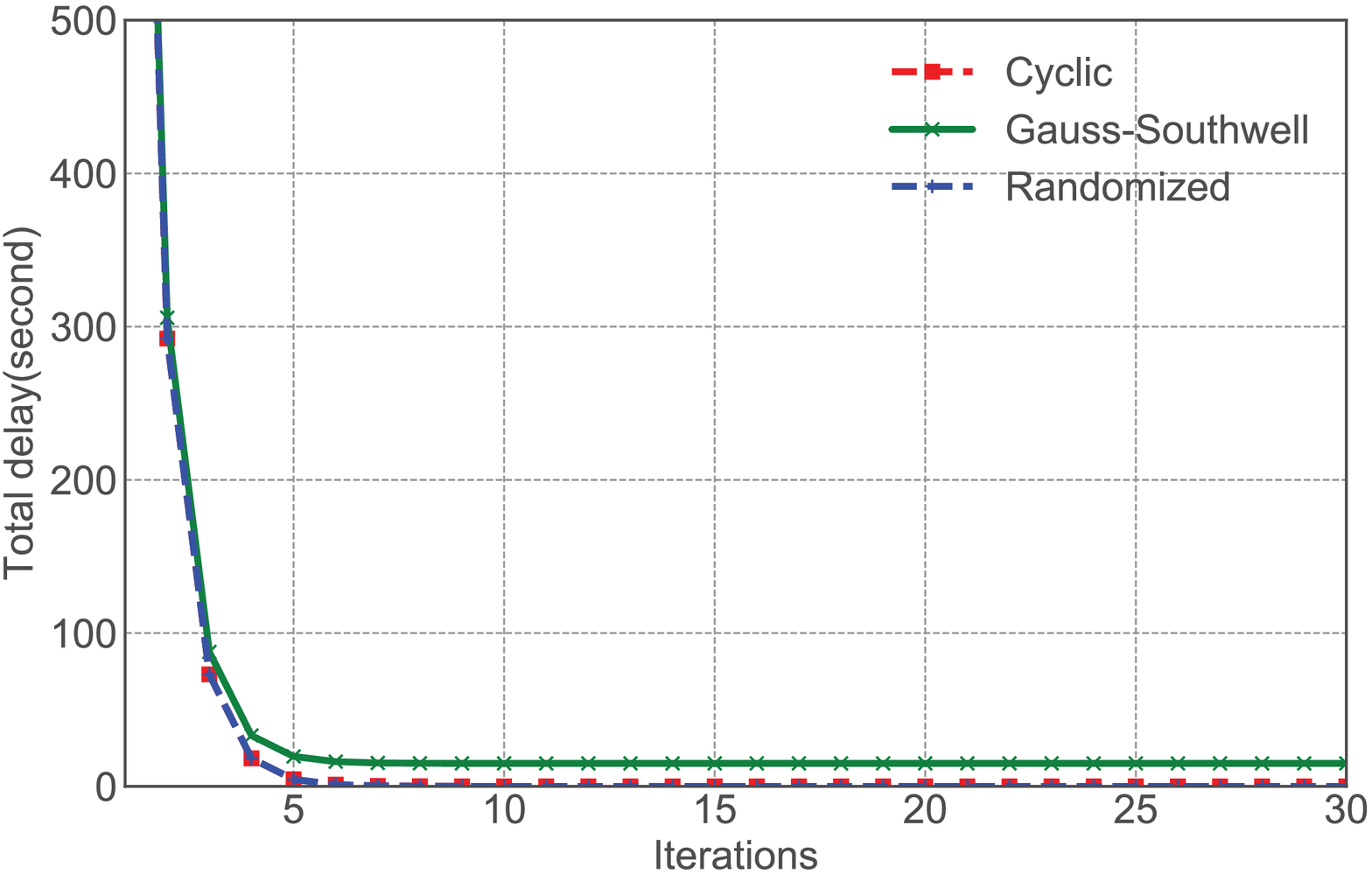}
		\caption{The solution of total delay minimization problem (\ref{eq:proximal_problem}).}
		\label{fig:total_delay_minimization}
	\end{minipage}

	\begin{minipage}{0.45\textwidth}	
		\centering
		\includegraphics[width=0.95\columnwidth]{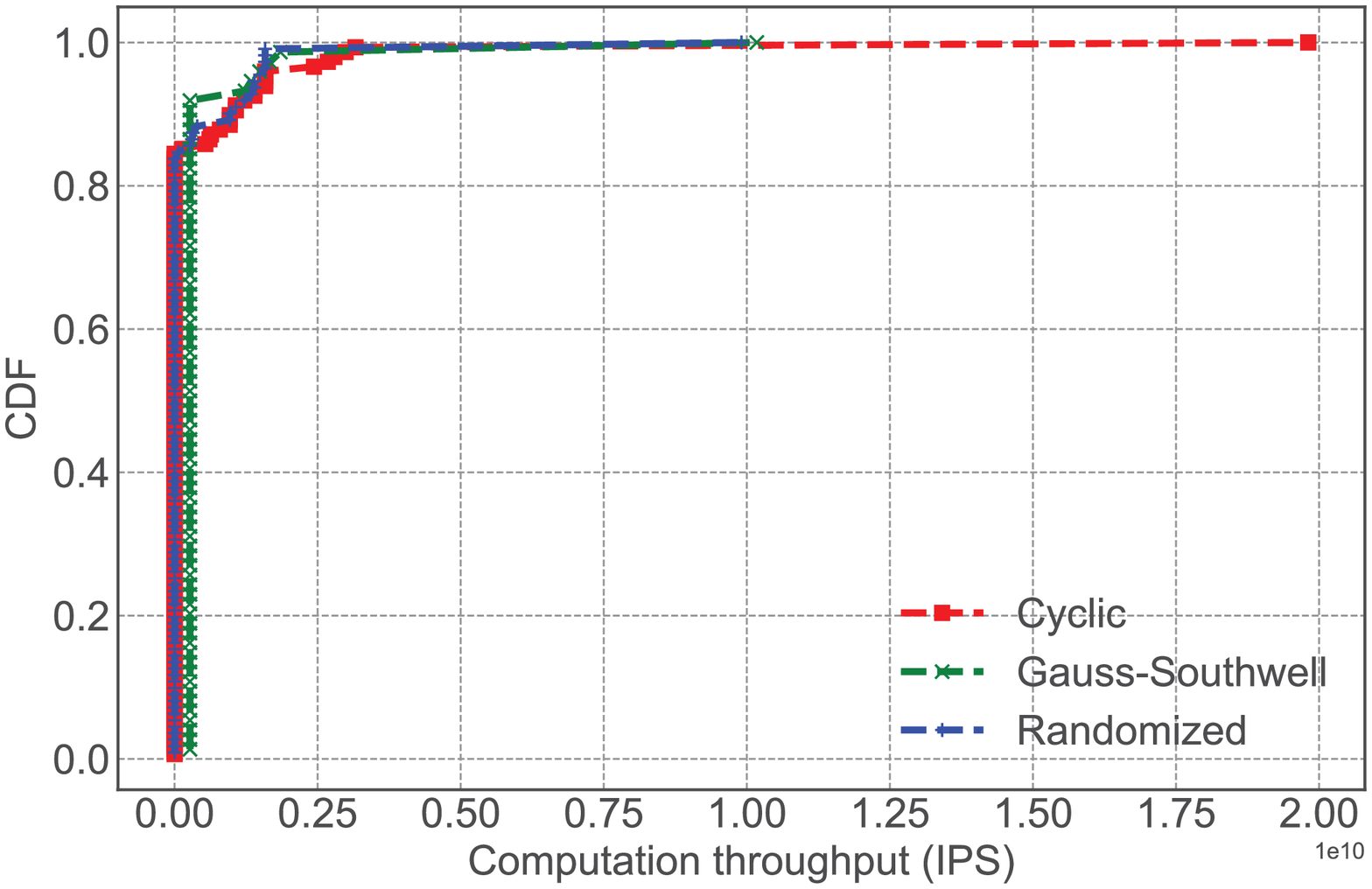}
		\caption{: Computation throughput for the cache contents.}
		\label{fig:computation}
	\end{minipage}
\end{figure}

For a self-driving car $v\in \mathcal{V}$, as shown in Fig \ref{fig:self_driving_passenger},  we generated randomly features of $37$ passengers (F: Female, M: Male, A: Anger, A*: Anticipation, D: Disgust, E: Joy, S*: Sad, S: Surprise, T: Trust). However, in a realistic implementation, for getting passengers' features, the CNN model described in Section \ref{subsubsec:CNN} should be used. For emotion-based clustering, we use $8$ emotion-based clusters:  anger, anticipation, disgust, fear, joy, sad, surprise, and trust as the labels. Furthermore, for age-based clustering, we use  $8$ age-based clusters: $[0\rightarrow9,10\rightarrow19,20\rightarrow29,30\rightarrow39,40\rightarrow49,50\rightarrow59,60\rightarrow69,70\rightarrow79]$ as the labels. We generated randomly demands  for contents and the popularity of the contents follows Zipf distribution described in \cite{newman2005power, ndikumana2017network}. Furthermore, the self-driving car has a WiFi bandwidth of  $160$ MHz (802.11ac) with a maximum theoretical data rate of $\tilde{\psi}^v_u=3466.8$  Mbps. In addition, the computation capacity of the self-driving car is set to $p_v=3.6 $ GHz, while the cache capacity is set to $c_v=100$ TB.

\subsection{Evaluation Results}
\label{subsec:ES}

Based on video ratings and users' location information, we select six areas to deploy RSUs by using the k-means algorithm. In the selected six areas, we predict the probabilities of contents to be requested in these areas by using MLP. As shown in Fig. \ref{fig:cache_probabilities}, in MLP, we minimize the cross-entropy loss function. An accuracy of $97.82\%$ is achieved for predicting the probabilities of contents to be requested in $6$ areas of RSUs. Each RSU $v\in \mathcal{V}$ caches movies by starting with the movies that have high ratings and predicted probabilities to be requested within the RSU area (in descending order) until the cache storage becomes full or there are no more movies to cache. As an example, Fig. \ref{fig:cache_probabilities} shows the top $8$ movies that need to be cached at RSU $1$ with their predicted probabilities using MLP.

Caching at the RSUs is based on location and movie ratings. However, in addition to location and movie ratings, caching in self-driving cars is based on passengers' features such as age, emotion, and gender. Therefore, when the self-driving car is connected to an RSU, it downloads the MLP output from the RSU. Then, it groups the MLP output based on age and emotion using the k-means algorithm and on gender using binary classification described in  Section  \ref{subsubsec:Recommendation}. Here, we use $8$ age-based clusters, $8$ emotion-based clusters, and $2$ gender-based clusters. Furthermore, for the passengers, we use age, emotion, and gender features described in Fig.  \ref{fig:self_driving_passenger}. However, CNN can be used to predict these features (age, emotion, and gender) using facial images of passengers captured by car's camera. The self-driving car uses k-means and binary classification to classify the passengers in different age, emotion, and gender-based clusters formed using MLP output. Then, inside the formed clusters,  the self-driving car finds the movies that have high ratings and predicted probabilities to be requested as recommended movies for the passengers. 

\textcolor{black}{Fig. \ref{fig:car_cache_recommendation} shows recommended movies to watch depending on age, emotion, and gender of the passengers. As shown in this figure, based on these features,  passengers may like similar movies (many passengers like Once Upon A Time and Secrets \& Lies). Therefore,  caching these recommended movies inside the car can prevent repetitive demands of the same movies that need to be sent to RSUs or DC. In other words, we can save bandwidth.} Furthermore, we chose CNN and MLP-based recommendation for movies over collaborative filtering approaches because each passenger’s features for infotainment contents are not a priori known by the self-driving car. The collaborative filtering approaches, which are described in \cite{subramaniyaswamy2017personalised}, consist of establishing the relationship between prior known users' preferences and movies' features. However, after identifying passengers' features and movies' features, we compare our proposal denoted Deep3C with the well-known collaborative filtering approaches such as Singular Value Decomposition (SVD), Non-negative Matrix Factorization (NMF), K-Nearest Neighbors (KNN), and  Co-clustering (Coclust).The simulation results in Fig. \ref{Algorithm_comparison} show that our proposal (Deep3C) achieves better performance over existing collaborative filtering approaches.

 We generated randomly demands of passengers for contents, where the popularity of the contents follows Zipf distribution \cite{newman2005power}. We use Zipf parameter $a$ with values from $a = 0.5$ to $a = 2.0$. The choice of a = 0.5 to a = 2.0 comes from the results presented in Fig. \ref{fig:content_distribution}, where the difference in convergence is observed within a range of $a = 0.5$ to $a = 2.0$. Furthermore, based on the demands of the passengers, Fig. \ref{fig:cache_hit_car} shows the normalized cache hits for the cached movies. The movies that are not cached in the self-driving car (cache misses) need to be retrieved at the RSU or DC. In this figure, we present the cache hits for the contents cached at RSUs and self-driving car. In other words, the total cache hits at RSUs and the self-driving car equal to $61\%$ of the whole demands, i.e., $39\%$ of the demands need to be served by DC. Therefore, with edge caching at RSUs and self-driving cars, we can significantly save backhaul bandwidth.   The results in this figure demonstrate that the cache hits increase with Zipf parameter, i.e., when $a = 2.0$  the small number of movies are very popular and requested by many passengers. In other words, the movies with high demands are characterized by high probabilities of being requested and caching these movies contribute to the high increase of cache hits.

Fig. \ref{fig:total_delay_minimization} shows the solution of the surrogate function (\ref{eq:proximal_problem}), where (\ref{eq:proximal_problem}) minimizes the total delays (transmission delay and computation delay).   The surrogate function (\ref{eq:proximal_problem}) converges to a coordinate-wise minimum point which is the stationary point through the use of different selection rules such as Cyclic, Gauss-Southwell, and Randomized. In other words, at a stationary point, the problem (\ref{eq:proximal_problem}) cannot find a better minimum direction. Furthermore, in this figure, the self-driving car needs to download the recommended contents first, and then caches these recommended contents; this contributes to high latency at the first iterations. As described in Fig \ref{fig:car_cache_recommendation}, some passengers may need to watch similar movies, i.e., many requests for movies can be satisfied from the cache storage.   

In Fig. \ref{fig:computation}, we present the Cumulative Distribution Function (CDF) of computational throughput in terms of the number of Instruction Per Second (IPS). Here, we define computation throughput as a measurement of how many units of tasks that can be computed by OBU for a given time. In this figure, the simulation results demonstrate that the Cyclic selection rule uses higher computational resource than Gauss-Southwell and Randomized selection rules. Cyclic selection rule has to choose index $j \in \mathcal{J}^t$ cyclically until all indexes in $\mathcal{J}^t$ are used.

\section{Conclusion}
\label{sec:Conclusion}
In this paper, we proposed a novel framework that uses deep learning for content caching in a self-driving car. In the proposed framework, at the DC, we proposed an MLP to predict the probabilities of contents being requested in specific areas. Then, the output is deployed in MEC servers (at the RSUs) close to the self-driving cars, where each MEC server downloads and caches the contents that have high probabilities of being requested in its coverage area. Furthermore, for a self-driving car, to cache infotainment contents that are appropriate regarding the age, emotion, and gender of the passengers,  we proposed to use CNN approach for predicting the age, emotion, and gender. Then, the self-driving car downloads the MLP output from the MEC server and combines CNN output with the MLP output using k-means and binary classifications to identify the infotainment contents that meet passengers' features to be downloaded and cached. Therefore, we formulated the deep learning-based caching problem as an optimization problem that minimizes the content-downloading delay. \textcolor{black}{ The simulation results demonstrate that our caching approach can reduce $61\%$ of the backhaul traffic, i.e., caching at RSUs and self-driving cars can serve $61\%$ of the whole demands for infotainment contents. Furthermore, our prediction for the infotainment contents that need to be cached at the  RSUs and the self-driving cars reaches  $97.82\%$ accuracy.}  
\bibliographystyle{IEEEtran}
\bibliography{references}
\vspace{2.0\baselineskip}

\end{document}